\documentclass[12pt]{article}
\usepackage{epsf}
\usepackage{amsmath}
\usepackage{graphicx}
\usepackage{color}
\usepackage{psfrag}

\usepackage{epsf}
\usepackage{graphicx,epsfig}
\usepackage{amssymb}
\usepackage{epstopdf}
\usepackage{amsmath}
\usepackage{amsfonts}
\usepackage{latexsym}
\usepackage{color}
\usepackage{xypic}
\usepackage{cancel,slashed}
\usepackage[linktocpage]{hyperref}
\usepackage{float}

\usepackage{ifpdf}

\def\IC{\mathbb{C}}

\def\IZ{\mathbb{Z}}

\def\IP{\mathbb{P}}

\def\s{{\text{sgn}}}

\def\tr{\mbox{Tr}}

\def\be{\begin{equation}}
\def\ee{\end{equation}}
\def\bea{\begin{eqnarray}}
\def\eea{\end{eqnarray}}
\def\bes{\begin{subequations}}
\def\ees{\end{subequations}}

\newcommand{\bmat}{\left(\begin{array}}
\newcommand{\emat}{\end{array}\right)}

\def\yzero{\smash{\hbox{$y\kern-4pt\raise1pt\hbox{${}^\circ$}$}}}

\def\beq{\begin{equation}}
\def\eeq{\end{equation}}
\def\beqa{\begin{eqnarray}}
\def\eeqa{\end{eqnarray}}

\def\-{\hphantom{-}}

\def\s2{\frac{1}{\sqrt2}}

\def\tr{{\rm tr \,}}

\def\IF{\relax{\rm I\kern-.18em F}}
\def\II{\relax{\rm I\kern-.18em I}}

\def\Dsl{\,\raise.15ex\hbox{/}\mkern-13.5mu D} %this one can be subscripted

\def\IC{{\bf C}}
\def\IS{{\bf S}}

\def\IZ{{\mathbb{Z}}}
\def\IX{{\bf X}}

\def\IP{\bf P}

\def\fref{Figure \ref}
\def\sref{\S \ref}

\catcode`\@=11   
\newdimen\@rotdimen
\newbox\@rotbox  

\def\@vspec#1{\special{ps:#1}}%  passes #1 verbatim to the output
\def\@rotstart#1{\@vspec{gsave currentpoint currentpoint translate
   #1 neg exch neg exch translate}}% #1 can be any origin-fixing transformation
\def\@rotfinish{\@vspec{currentpoint grestore moveto}}% gets back in synch 
\def\@rotr#1{\@rotdimen=\ht#1\advance\@rotdimen by\dp#1%
   \hbox to\@rotdimen{\hskip\ht#1\vbox to\wd#1{\@rotstart{90 rotate}%
   \box#1\vss}\hss}\@rotfinish}
\def\@rotl#1{\@rotdimen=\ht#1\advance\@rotdimen by\dp#1%
   \hbox to\@rotdimen{\vbox to\wd#1{\vskip\wd#1\@rotstart{270 rotate}%
   \box#1\vss}\hss}\@rotfinish}%
\def\@rotu#1{\@rotdimen=\ht#1\advance\@rotdimen by\dp#1%
   \hbox to\wd#1{\hskip\wd#1\vbox to\@rotdimen{\vskip\@rotdimen
   \@rotstart{-1 dup scale}\box#1\vss}\hss}\@rotfinish}%
\def\@rotf#1{\hbox to\wd#1{\hskip\wd#1\@rotstart{-1 1 scale}%
   \box#1\hss}\@rotfinish}%
\def\rotate{\@ifnextchar[{\@rotate}{\@rotate[l]}}
\def\@rotate[#1]#2{\setbox\@rotbox=\hbox{#2}\@nameuse{@rot#1}\@rotbox}

\catcode`\@=12
\topmargin
-1.5cm
\textwidth
15.5cm
\textheight
23.5cm
\oddsidemargin
0.7cm
\evensidemargin
1.2cm

\begin{document}

%----------------------------------------------------------------------%
%  numbering equations with section number
%----------------------------------------------------------------------%
\makeatletter
\@addtoreset{equation}{section}
\makeatother
\renewcommand{\theequation}{\thesection.\arabic{equation}}
\renewcommand{\thefootnote}{\fnsymbol{footnote}}
%----------------------------------------------------------------------%
%  title page
%----------------------------------------------------------------------%
\pagestyle{empty}
%\vspace*{1.0in}
\rightline{IFT-UAM/CSIC-14-046, FTUAM-14-19}
\rightline{IPPP/14/49, DCPT/14/98}
\vspace{0.1cm}
\begin{center}
\LARGE{\bf Axion Monodromy Inflation \\ on Warped Throats \\[12mm]}
\large{Sebasti\'an Franco$^1$\footnote{sebastian.franco@durham.ac.uk}, Daniele Galloni$^{1}$\footnote{daniele.galloni@durham.ac.uk},\\ Ander Retolaza$^{2,3}$\footnote{ander.retolaza@uam.es},  Angel Uranga$^3$\footnote{angel.uranga@uam.es}\\[3mm]}
\footnotesize{
$^1$ Institute for Particle Physics Phenomenology, Department of Physics,\\[-0.3em] 
Durham University, Durham DH1 3LE, United Kingdom \\[2mm]
${ }^2$ Departamento de Fisica Te\'orica, Facultad de Ciencias \\[-0.3em] 
Universidad Aut\'onoma de Madrid, 28049 Madrid, Spain\\[2mm]
$^3$ Instituto de Fisica Te\'orica IFT-UAM/CSIC,\\[-0.3em] 
C/ Nicol\'as Cabrera 13-15, Universidad Aut\'onoma de Madrid, 28049 Madrid, Spain } \\[2mm] 
\small{\bf Abstract} \\[5mm]
\end{center}
\begin{center}
\begin{minipage}[h]{16.0cm}
Recent models of axion monodromy inflation in string theory link the inflationary potential and the moduli stabilization potential. Realistic inflationary models require mechanisms to moderately suppress the inflaton mass with respect to the moduli stabilization scale. In this paper we explore the realization of this idea using warped throats, whose redshifted infrared region supports the inflaton mode. The inflaton potential and its monodromy arise from couplings to the fluxes supporting the throat. We provide explicit realizations of such throats in type IIB with NSNS and RR 3-form field strength fluxes, and in type IIA with RR 2-form fluxes. Once embedded in a global CY, these systems would provide a mechanism to realize chaotic inflation at scales parametrically suppressed with respect to bulk physics. The construction of the throats is systematically carried out using geometric transitions in systems of D-branes at singularities, whose properties and dynamics are efficiently encoded using dimer diagrams. The holographic dual of the axion monodromy is a quasi-periodic chain of Seiberg dualities.
\end{minipage}
\end{center}
\newpage
%----------------------------------------------------------------------%
%  Resetting of counters
%----------------------------------------------------------------------%
\setcounter{page}{1}
\pagestyle{plain}
\renewcommand{\thefootnote}{\arabic{footnote}}
\setcounter{footnote}{0}
%----------------------------------------------------------------------%
%  Paper begins
%----------------------------------------------------------------------%

\vspace*{1cm}

\setcounter{tocdepth}{2}

%\vspace*{-1cm}
\tableofcontents

\vspace*{1cm}

%=======================================================
\section{Introduction and review of axion monodromy}
%=======================================================
%

The underlying continuous shift symmetry of axions makes these fields good candidates for the inflaton, once a controllable potential (breaking the continuous symmetry but preserving a discrete periodicity) is introduced. In order for axions with sub-Planckian periods to achieve super-Planckian field ranges (as seemingly favored by the recent BICEP2 data [1]), it is natural consider a monodromic potential, i.e. a multivalued function of the axion, preserving the discrete periodicity yet allowing for super-Planckian excursions \cite{am1,am2}. \footnote{Other approaches, based on multiple and/or aligned axions, have been considered in e.g. \cite{Dimopoulos:2005ac,Kim:2004rp,Kappl:2014lra,Berg:2009tg}.}

Axions are ubiquitous in string theory, as they arise in the KK compactification of higher dimensional $p$-form gauge potentials. The continuous shift symmetry is inherited from the higher dimensional gauge invariance, and although it is broken by non-perturbative effects from euclidean string or brane instantons, charge quantization of the latter allows a discrete periodicity to survive. 

In early axion monodromy inflation models \cite{am1,am2}, the axion arises from e.g. the RR 2-form on 2-cycles, and the monodromy is achieved by introducing NS5-branes (actually, brane-antibrane pairs, due to tadpole cancellation in compact examples) wrapped on the 2-cycle, such that the shift of the axion produces an energy increase due to the induced D3-brane charge. In order to suppress brane-antibrane annihilation, or to keep backreaction under control, the systems are proposed to be located down warped throats \cite{Conlon:2011qp,Flauger:2009ab}. Related variants have been discussed in \cite{Palti:2014kza}. These systems are inherently strongly coupled, and are actually constructed as the S-duals of models with an axion from the NSNS 2-form and monodromy from D5-branes.

A new and better monodromy inflation framework was proposed in \cite{Marchesano:2014mla} (see also \cite{Blumenhagen:2014gta,Hebecker:2014eua, Ibanez:2014kia,Arends:2014qca,McAllister:2014mpa} for subsequent work), in particular in flux compactifications, based on a topological effect of fluxes anticipated in \cite{BerasaluceGonzalez:2012zn}. For our purposes, the key ingredient in fluxed axion monodromy arises from the 4d couplings in the KK reduction of the Chern-Simons (CS) terms in the 10d action, in the presence of fluxes. For instance, consider the 10d coupling in type II strings (IIA/B for $p$ even/odd)
\beqa
\int_{10d} B_2 \wedge F_p\wedge F_{8-p}
\label{cs-general}
\eeqa
and compactify to 4d on a (not necessarily Calabi-Yau) space $\IX_6$, with a 2-cycle $\Sigma_2$ and a $p$-cycle $\Pi_p$. We define the flux and 4d fields
\beqa
\int_{\Pi_p} F_p=M\quad ,\quad \phi=\int_{\Sigma_2} B_2\quad ,\quad F_4=\int_{\Pi'_{4-p}} F_{8-p}
\label{flux-general}
\eeqa
where $\Pi'_{4-p}$ is transverse to $\Sigma_2$ and $\Pi_{p}$ in $\IX_6$ (concretely, the dual of $\Sigma_2 \times \Pi_p$). As explained in  \cite{Marchesano:2014mla}, the 4d axion $\phi$ has a monodromy induced by the flux $F_p$. Due to the 10d CS term, the physical  RR $(p+2)$-form field strength  is
\beqa
\tilde{F}_{p+2}= F_{p+2}+ F_p \wedge B_2
\label{theflux}
\eeqa
with $F_{p+2}=dC_{p+1}$. Hence, a change in $\phi$ away from its minimum increases the  physical $(p+2)$-form flux on 
$\Sigma_2 \times \Pi_p$. This leads to an inflaton potential, arising from the kinetic term for  $\tilde{F}_{p+2}$, and is hence quadratic at lowest order. Despite the monodromy, the system has an underlying periodic structure because the theory contains 4d domain walls, given by D$(6-p)$-branes wrapped on $\Pi'_{4-p}$, which can change the $F_{p+2}$ background (and thus ${\tilde F}_{p+2}$) by an integer amount, thereby interpolating among the different branches.

In the 4d theory, we get a coupling
\beqa
M\,\int_{4d} \phi\, F_4 .
\eeqa
This 4d description is the one proposed in \cite{Kaloper:2008fb,Kaloper:2011jz,Kaloper:2014zba} as an effective action for axion monodromy. As emphasized there, and in  \cite{Marchesano:2014mla}, the axion shift symmetry is related to gauge invariance of a dual 3-form, in a generalization of the St\"uckelberg mechanism for 3-forms (see \cite{Berasaluce-Gonzalez:2013bba} for related discussions, and \cite{Dvali:2005ws,Dvali:2005an,Dvali:2013cpa} for this mechanism applied to the QCD axion).

In this framework, the appearance of the monodromy is related to the fluxes stabilizing moduli in the model. The intricate relation between the inflationary potential and the moduli stabilization potential is an interesting feature of these models, as emphasized in \cite{Marchesano:2014mla,McAllister:2014mpa}. However, realistic applications to inflationary models complying with the recent BICEP2 results demand a moderate hierarchy between the moduli stabilization scale, and the inflation scale $\sim 10^{16}$ GeV (or perhaps even the inflaton mass $\sim 10^{14}$ GeV). Although this point has been recognized in the literature, it has not been properly addressed hitherto. In this paper we consider the explanation of this hierarchy by using warped throats.

We provide an explicit construction of local warped throats in type IIB with moduli stabilization by 3-form fluxes (generalizing \cite{Klebanov:2000hb}), and whose infrared region supports an axion (arising from 2-cycles at the tip of the throat) with a monodromy induced by the 3-form flux itself. The geometry of the throats is based on performing a geometric transition in a systems of fractional D3-branes at a toric singularity, in which some 2- and 4-cycles shrink and are replaced by 3-cycles which support RR 3-form flux. The properties of the throat are nicely encoded in the holographic dual quiver field theory (describing the D-branes before the geometric transition), which is described by a dimer diagram. We provide several explicit examples, and present techniques to construct fairly general classes of such throats. We also present an analogous construction for type IIA throats, which are obtained from (the inverse) geometric transitions in systems of D6-branes on 3-cycles, which shrink and are replaced by 2-cycles supporting RR 2-form flux. The warped throats we describe in this paper are amenable to embedding in global compactifications (fairly easily in the case of type IIA models, and in type IIB if one allows for global non-trivial 1-cycles), although such global embeddings are not explicitly discussed and are left for future work.

The paper is organized as follows. In section \ref{sec:throats} we describe the construction of our warped throats. In section \ref{sec:ks} we review the Klebanov-Strassler throat for the conifold \cite{Klebanov:2000hb}, in section \ref{sec:monodromy} we describe the appearance of axions and monodromy when the throat includes a 2-cycle at its tip. In section \ref{section_dP3} we focus on a very explicit case study, a deformation of the complex cone over the del Pezzo surface $dP_3$. In sections \ref{section_KT_dP3} and \ref{section_bottom_throat_dP3} we describe the throat and the 3- and 2-cycle structure, both from the geometry and the holographic field theory dual. In section \ref{sec:seiberg} we discuss the description of the axion monodromy in terms of the field theory dual, which turns out to correspond to a cascade of Seiberg dualities. 
In section \ref{sec:iia} we describe the type IIA implementation of these ideas, which involves only 2-cycles. In section  \ref{sec:implications} we discuss implications for inflationary models. Finally in section \ref{sec:conclu} we present our conclusions. In appendix \ref{sec:general} we exploit toric geometry and dimer technology to construct a general class of throats with one 3-cycle supported by fluxes, and one axion with monodromy, and discuss the gauge dynamics underlying the geometric transition to the deformed geometry. In appendix \ref{sec:noint} we discuss a subtle example, an orbifold of the conifold, in which there is no axion monodromy due to the absence of a certain geometric intersection number in the geometry.

\bigskip

%=======================================================
\section{Warped Throats}
%=======================================================
\label{sec:throats}

%=======================================================
\subsection{Review of the KS Throat}
%=======================================================
\label{sec:ks}

Warped throats have become  a standard tool to generate hierarchies, which moreover admits an interpretation in terms of dimensional transmutation via the gauge/gravity duality. The prototypical example is the Klebanov-Strassler (KS) throat \cite{Klebanov:2000hb}, which is obtained  by considering the deformed conifold 
\beqa
xy-zw=\epsilon
\label{conifold}
\eeqa
and introducing $M$ units of RR 3-form flux $F_3$ on its $\IS^3$,\footnote{The $\IS^3$ is manifest by changing the equation to $\sum_{i=1}^4 x_i^{\,2}=|\epsilon|$, and restricting to real $x_i$.} along with a non-trivial NSNS 3-form background $H_3$ in the dual 3-cycle, which is non-compact in the present local model. The complete supergravity solution is explicitly known for this simple case, and falls in the framework of type IIB flux compactifications with imaginary self-dual 3-form flux $G_3=F_3-\tau H_3$ \cite{Dasgupta:1999ss,Giddings:2001yu}. The flux stabilizes the modulus (via the flux superpotential in \cite{Gukov:1999ya}) and redshifts scales by a warp factor as follows
\beqa
\epsilon \sim \exp \Big(-\frac{2\pi K}{Mg_s}\;\Big)\quad, \quad m\sim \exp\Big( -\frac{2\pi K}{3Mg_s}\;\Big)
\label{ks-scale}
\eeqa
where $K$ denotes the NSNS 3-form flux quantum, after cutting off the throat at some distance in the radial direction, equivalently an energy scale in the dual theory. The radial direction of a throat is holographically interpreted as the energy scale in the dual gauge theory. For this reason, we will often use the common terminology of UV and IR to refer to the large and small radius regions, respectively. The main lesson of our previous discussion is that the dynamics down the throat is exponentially suppressed with respect to the UV scale in the bulk of the compactification.

The above picture generalizes to other Calabi-Yau (CY) singularities admitting a complex deformation. There are warped throat supergravity solutions with RR 3-form flux on the 3-cycles of the deformation and NSNS 3-form flux on the non-compact duals, with the warped metric being conformal to the underlying deformed CY metric. General throat supergravity solutions simplify in the $r\gg \epsilon$ regime, where their metric is a warped version of the underlying conical singularity metric. The classic example of such geometry is the $r\gg \epsilon$ regime of KS, also known as the Klebanov-Tseytlin (KT) solution \cite{Klebanov:2000nc}.

\bigskip

%=======================================================
\subsection{Warped Axion Monodromy}
%=======================================================
\label{sec:monodromy}

The idea now is to apply the warp suppression mechanism to axion monodromy inflation. We would like to consider a warped throat, based on a complex deformation of a CY singularity given by a real cone over a Sasaki-Einstein 5-manifold $\IX_5$. We consider the geometry of the throat to contain a non-trivial 2-cycle $\Sigma_2$, for simplicity an $\IS^2$, at its bottom. The throat will be supported by a 3-cycle $\Pi_3$, with $M$ units of RR 3-form flux. An explicit example will be discussed in coming sections. Application of (\ref{cs-general}) and (\ref{flux-general}) to this case leads to
\beqa
\int_{10d} F_3\wedge B_2\wedge F_5 \quad \to \quad M\int_{4d}  \phi \, F_4
\eeqa
with $\phi$ the integral of the NSNS 2-form over $\Sigma_2$, and $F_4$ the integral of $F_5$ along the radial direction. As usual, this is a bit ill-defined in the non-compact setup, and should be more properly regarded by cutting off the throat at some scale (or including a compactification).
For the above KK reduction to produce the 4d topological term, we need a non-trivial wedge produc of $F_3$ and $B_2$, equivalently, there must be a non-trivial intersection $q=[\Pi_3]\cdot[\Sigma_2]$ in $\IX_5$.

This system has an axion monodromy for $\phi$, which increases the flux ${\tilde F}_5\sim F_3\wedge B_2$ along $\IX_5$ (by $qM$ units). The associated domain wall is given by a D3-brane stretching in the radial direction. As in \cite{BerasaluceGonzalez:2012zn}, the domain wall is $\IZ_M$ valued, as $M$ such domain walls can decay (by ending on a string, given by a NS5-brane on the $\IS^3$ times the radial direction, which has a Freed-Witten anomaly and therefore must spit off D3-brane domain walls). Note that these domains walls are different from those in the literature e.g. \cite{Gubser:1998fp}.

\medskip

Clearly, there are many other similar systems that can be constructed. For instance, the S-dual configurations of the above systems provide a realization of axion monodromy in which the axion arises from the RR 2-form, rather than the NSNS 2-form. This may be useful for applications to inflation in compactifications in which the NSNS axion suffers from eta problems. In fact, these throats can be regarded as holographic duals of NS5-brane models similar to those considered in \cite{am1,am2}; namely, the NS5-branes on 2-cycles are replaced by NSNS $H_3$ fluxes on the 3-cycle after the geometric transition. This representation has the advantage of admitting a description in string perturbation theory.

For concreteness we stick to our original realization in terms of an NSNS axion and RR fluxes (and the RR axion models can be obtained by a mere S-duality). A related, but interesting in itself, realization in type IIA models is discussed in section \ref{sec:iia}. In the next section we present a detailed analysis of a particular example based on a complex cone over $dP_3$. The general discussion is subsequently resumed, and can be followed fairly independently.

\bigskip

%=======================================================
\section{An Explicit Example Based on Del Pezzo 3}
%=======================================================

\label{section_dP3}

It is easy to cook up throats with both 2- and 3-cycles at their bottom. Throughout this article, we will focus on models arising from toric Calabi-Yau 3-folds. Restricting to this class of models has the advantage that several details of the geometry, including its complex deformations, can be succinctly captured by $(p,q)$ web diagrams \cite{Aharony:1997ju,Aharony:1997bh}. In addition, dimer models provide a powerful tool for connecting them to the corresponding holographic dual quantum field theories and studying their dynamics. Dimers are by now standard tools, we refer the interested reader to \cite{Franco:2005rj, Kennaway:2007tq, Yamazaki:2008bt} for detailed references. Although many of the results below are derived in the literature, we put them to work to develop a clear picture of the warped axion monodromy construction.

We will now introduce an explicit realization of the warped monodromy scenario based on the complex cone over $dP_3$. Over the next couple of sections, we will discuss it from both a geometric and field theoretic perspectives. Complex cones over del Pezzo surfaces have been extensively studied in the context of the gauge/gravity correspondence for D3-branes at singularities \cite{Beasley:1999uz}-\cite{Feng:2002zw}.  A classification of more general geometries suitable for warped monodromy is postponed to Appendix \ref{sec:general}.

The del Pezzo surface $dP_n$ can be constructed as $\IP_2$ blown up at $n$ generic points. It contains $n+1$ 2-cycles, denoted $H$ (inherited from the $\IP_1\subset \IP_2$), and $E_I$, $I=1,\ldots, n$ (the $n$ blown-up 2-cycles). One can build a 5-manifold admitting a Sasaki-Einstein metric by fibering an $\IS^1$ over $dP_n$, with Chern class given the K\"ahler class of $dP_n$. Finally, one can construct a conical CY singularity $\IX_6$ as a real cone over $\IX_5$; this has the structure of a complex cone over $dP_n$. For $n\leq 3$, the singularities can be described using toric geometry, which reduces the complex geometry of these spaces to simple diagrams.

\begin{figure}[!ht]
\begin{center}
\includegraphics[width=4.5cm]{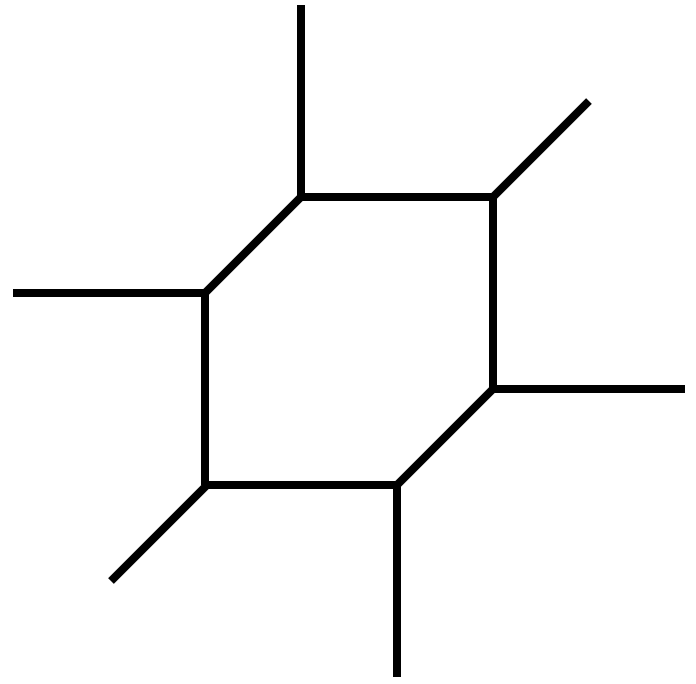}
\caption{$(p,q)$ web diagram for the complex cone over $dP_3$. For clarity we show the 4- and 2-cycles as slightly blown up. }
\label{web_dP3}
\end{center}
\end{figure}

We focus on the complex cone over $dP_3$ which is a toric CY 3-fold, whose geometry can be nicely encoded in the $(p,q)$ web diagram shown in \fref{web_dP3}. 

%=======================================================
%=======================================================

\begin{figure}[!ht]
\begin{center}
\includegraphics[width=13cm]{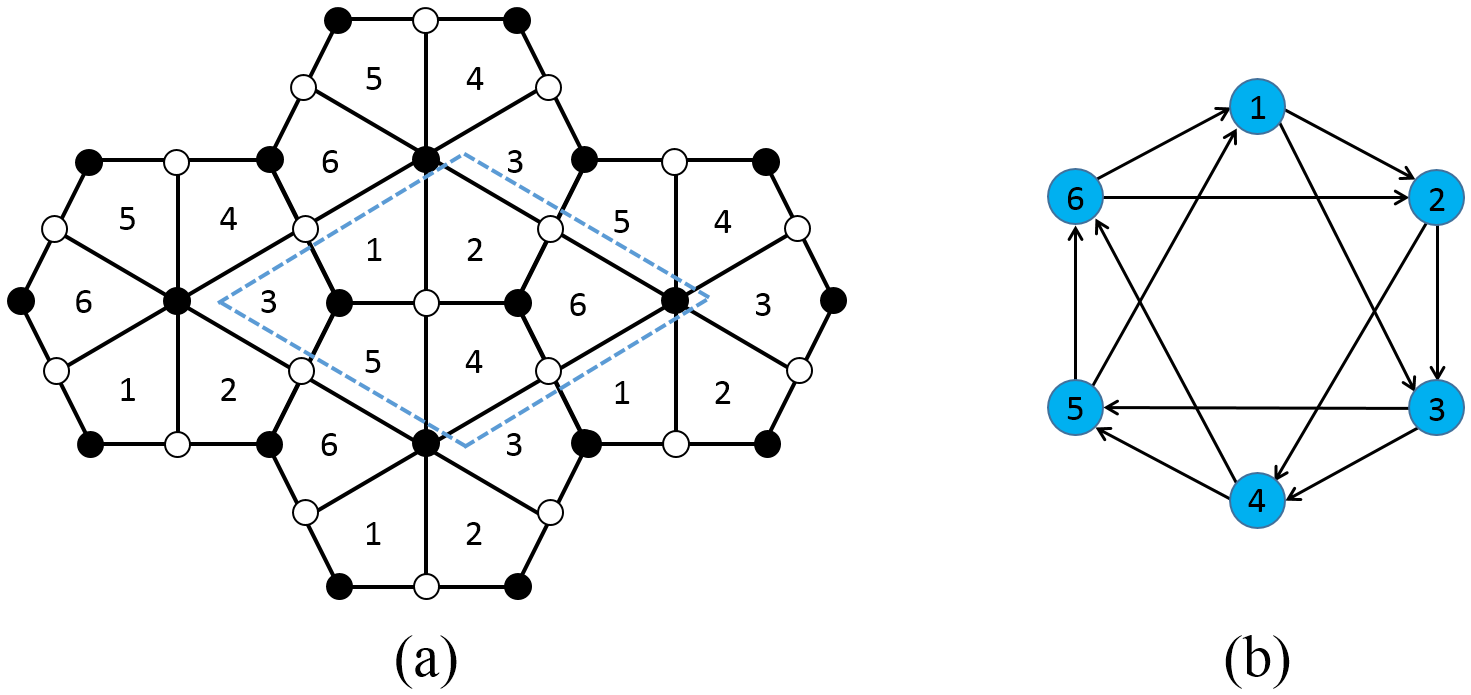}
\caption{a) Dimer model for phase 1 of $dP_3$. b) The corresponding quiver diagram.}
\label{dimer_quiver_dP3}
\end{center}
\end{figure}

The 4d $\mathcal{N}=1$ gauge theory on D3-brane probing this cone is nicely encoded by the dimer diagram in \fref{dimer_quiver_dP3}.a. The quiver diagram for this theory is given in \fref{dimer_quiver_dP3}.b. In fact, this theory is only one of four so-called {\it toric phases} interconnected by Seiberg duality, which is often referred to as phase 1 \cite{Feng:2001xr}-\cite{Feng:2002zw}.

%=======================================================
%=======================================================

Let us anticipate how this model contains some of the main ingredients of warped axion monodromy. As we will discuss in \sref{section_bottom_throat_dP3}, this geometry admits a complex deformation in which a finite size 3-cycle with $\IS^3$ topology appears. Furthermore, there will be a remnant singularity containing a vanishing $\IS^2$. Introducing $M$ units of RR 3-form flux on the 3-cycle and NSNS 3-form flux on its non-compact dual leads to a warped throat with the $\IS^2$ at its bottom that generalizes the KS throat.

Having to work with collapsed 2-cycles might not seem completely appealing. It is in principle possible to consider the blow-up of the 2-cycle, along the lines of \cite{Klebanov:2007us}, although this usually complicates the supergravity solution beyond tractability. Nevertheless, string theory is perfectly well-defined even in the limit of collapsed 2-cycle, since the non-trivial NSNS 2-form on it makes the physics smooth.

\bigskip

%=======================================================
\subsection{The KT Throat and its Dual RG Cascade}
%=======================================================

\label{section_KT_dP3}

Let us first consider throats in the KT regime, i.e. far from their IR bottom. In \cite{Franco:2004jz}, general warped geometries were constructed in this approximation. They are given by a warped product between Minkowski space and the CY cone $\IX_6$  of the form
\beq
ds^2=Z^{-1/2}\eta_{\mu\nu} dx^\mu dx^\nu + Z^{1/2} ds_X^2 .
\label{metric_throat}
\eeq
There are RR 3-form fluxes on the 3-cycles obtained as the $\IS_1$ fibration over the 2-cycles $E_i$ (in $\IX_5$), and NSNS 3-form fluxes on its duals. For $C_0=0$, we have the imaginary self-dual combination
\beq
G_3=F_3-{i\over g_s} H_3=\sum_{I} a^I \left(\eta + i {dr\over r} \right) \wedge \Phi_I,
\label{G3_throat}
\eeq
where $\Phi_I$ are $n$ harmonic $(1,1) $ forms associated to $E_I$ are $\eta$ is a 1-form along the $\IS^1$ fiber. Also $a^I=6\pi \alpha' M^I$ where $M_I$ correspond to the RR 3-form flux quanta. These fluxes source a warp factor 
\beq
Z(r) = {2 \cdot 3^4 \over 9-n} \alpha'^2 g_s^2 \left({\ln (r/r_0) \over r^4} + {1\over 4r^4}\right) \sum_{I,J}M^I A_{IJ}M^J ,
\label{warp_factor_throat}
\eeq
where $A_{IJ}$ is the associated intersection matrix.

Let us turn our attention to the dual field theory, for the $dP_3$ case. When all nodes in the quiver in \fref{dimer_quiver_dP3} have equal ranks $N_i\equiv N$, $i=1,\ldots,6$, the field theory is conformal and its gravitational dual is the space AdS$_5\times \IX_5$, with $N$ units of RR 5-form flux on $\IX_5$. This theory has marginal operators corresponding to modifications of the gauge couplings. Their gravity duals are the type IIB axio-dilaton (fixing the overall gauge coupling), and the integrals of the NSNS 2-form field $B_2$ over different 2-cycles in $\IX_5$, which are all moduli with no potential.

We are rather interested in the field theory dual to warped throats with fluxes. As mentioned above, this is achieved by introducing fractional branes, namely an anomaly-free change in the gauge factor ranks, corresponding to additional D5-branes wrapped on collapsed 2-cycles at the singularity.\footnote{More precisely, fractional branes are constrained by local tadpole cancellation, which is more restrictive than anomaly cancellation in the gauge theory. It corresponds to ``cancellation of non-abelian anomalies'' (i.e. the equality of incoming and outgoing arrows, counted with multiplicity) even for nodes of the quivers with ranks $N_a=0,1,2$ (see \cite{Leigh:1998hj,Aldazabal:1999nu} for early discussions in orbifolds).}

For concreteness, let us consider a single type of fractional branes, corresponding to the rank vector $\vec{N}=N(1,1,1,1,1,1)+M(1,0,0,1,0,0)$. Here $N$ and $M$ are the number of regular and fractional D3-branes, respectively. The effect of the fractional branes is to break conformal invariance. In the $M\ll N$ limit, it is possible to compute exact anomalous dimensions in the CFT limit and use them to determine beta functions.\footnote{This is true only if anomalous dimensions get corrections with respect to the conformal values at $\mathcal{O}(M^2/N^2)$. In some theories, such as the conifold, it is possible to show that this is the case using symmetry arguments in the gauge theory. In a large class of examples appearing in the literature, e.g. \cite{Franco:2004jz}, this behavior is simply assumed and then verified to be consistent with the dual supergravity solutions.}

As in \cite{Klebanov:2000hb}, this theory has a non-trivial RG flow that takes the form of a duality cascade, in which gauge groups are Seiberg dualized every time they reach infinite coupling. In preparation for our discussion in coming sections, we will refer to such a sequence of dualities as an {\it RG cascade}. 

%=======================================================
\begin{figure}[!ht]
\begin{center}
\includegraphics[width=12.5cm]{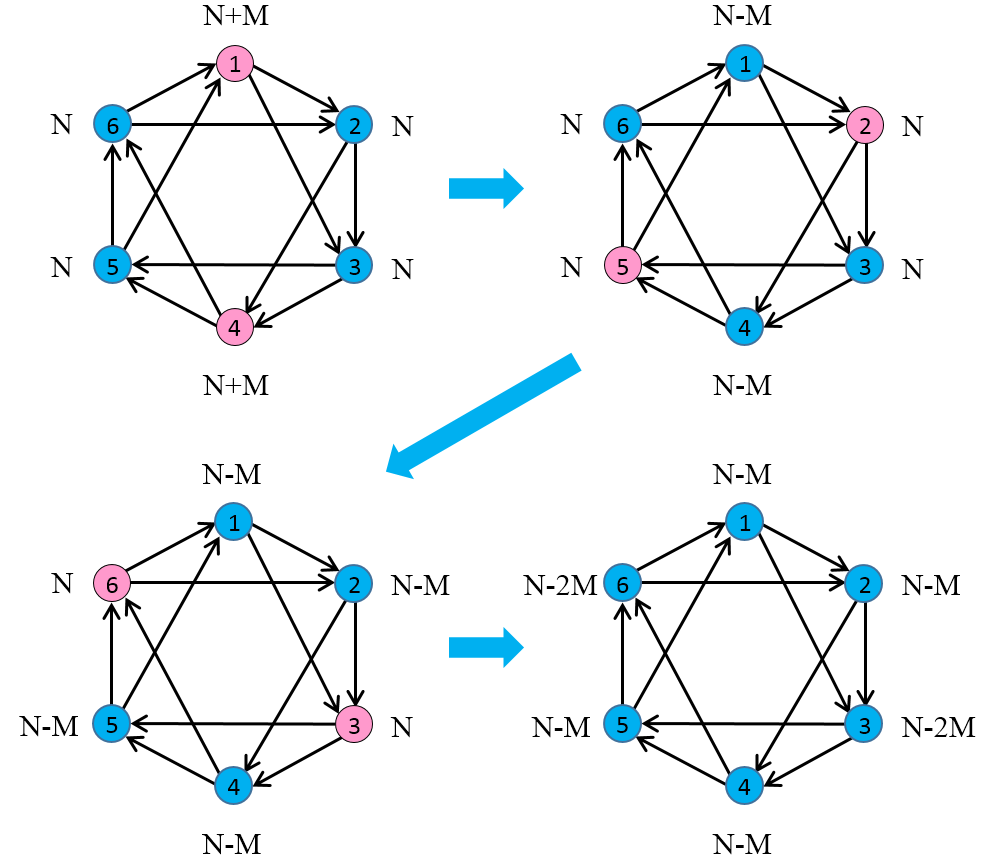}
\caption{A few steps in the RG cascade for $dP_3$ starting with ranks $\vec{N}=N(1,1,1,1,1,1)+M(1,0,0,1,0,0)$. The pair of nodes dualized at each time is shown in pink.}
\label{cascade_quiver_dP3}
\end{center}
\end{figure}
%=======================================================

In the example at hand, the RG cascade iterates the following sequence of dualization of antipodal pairs of nodes: (1,4), (2,5), (3,6) \cite{Franco:2005fd}. As shown in \fref{cascade_quiver_dP3}, after six dualizations we obtain the original quiver, but with ranks $\vec{N}=(N-M)(1,1,1,1,1,1)-M(0,0,1,0,0,1)$. Up to a trivial permutation of the nodes, we see that the number of regular D3-branes decreases according to $N\to N-M$ and the number of fractional branes transforms as $M\to -M$. In fact after an additional six dualizations we observe that the sign flip in the number of fractional branes is reversed, and we conclude that their number remains constant throughout the cascade.

This RG cascade is the field theory dual of a throat solution described by equations (\ref{metric_throat})-(\ref{warp_factor_throat}). 

\bigskip

%=======================================================
\subsection{The Bottom of the Throat: Complex Deformation from Strong Dynamics}
%=======================================================

\label{section_bottom_throat_dP3}

As explained in detail in \cite{Franco:2005fd}, complex deformations of toric singularities can be efficiently described in terms of $(p,q)$ webs. They translate to decomposing a $(p,q)$ web into sub-webs in equilibrium. This process is often referred to as geometric transition.

\begin{figure}[!ht]
\begin{center}
\includegraphics[width=9.5cm]{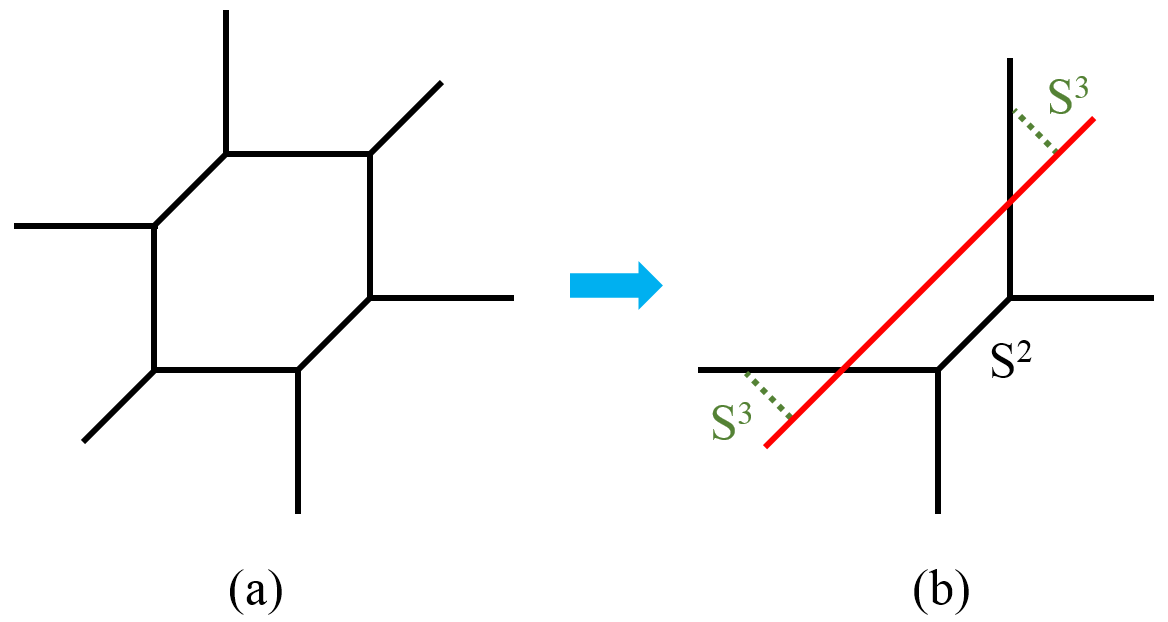}
\caption{a) Web diagram of the complex cone over $dP_3$. b) Complex deformation showing the 2- and 3-cycles in the resulting geometry. The two $\IS^3$, indicated by dashed lines, are actually homologous.
\label{fig:dp3}}
\end{center}
\end{figure}

\fref{fig:dp3} presents the particular deformation of the $dP_3$ singularity we are interested in.\footnote{Other deformations are possible. They are either equivalent by a rotation of the diagram, or do not include 2-cycles in the deformed geometry} As desired for axion monodromy applications, the geometry grows a finite size $\IS^3$, and in contrast with \cite{Klebanov:2000hb}, the deformed space is not completely smooth but possesses a conifold singularity, which contains a vanishing $\IS^2$. This geometry is the one at the bottom of the throat generated by the fractional branes considered in \sref{section_KT_dP3}. According to the holographic dictionary, this behavior should be  recovered from the IR dynamics of the dual gauge theory. We now review the basic ideas for understanding the deformation from gauge theory. A detailed discussion of this model can be found in \cite{Franco:2005fd}.

%=======================================================
%=======================================================

\begin{figure}[!ht]
\begin{center}
\includegraphics[width=11cm]{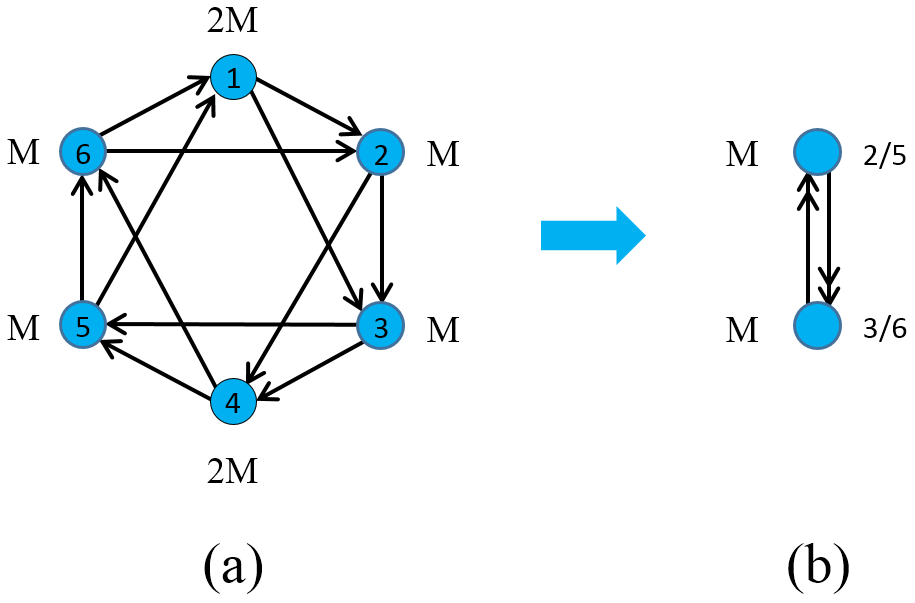}
\caption{a) The $dP_3$ quiver at the last step of the cascade. b) The conifold theory arises after higgsing by mesons vevs.}
\label{quiver_dP3_to_conifold}
\end{center}
\end{figure}

The RG cascade discussed in the previous section progressively reduces the number of regular D3-branes until reaching a point in the IR at which the ranks of the quiver become $\vec{N}=(2M,M,M,2M,M,M)$, as in \fref{quiver_dP3_to_conifold}.a. Nodes 1 and 4 have $N_f=N_c=2M$ and hence lead to a {\it quantum modified moduli space}. 

%=======================================================
%=======================================================

Every $N_f=N_c$ gauge group confines and gives rise to gauge invariant (from the point of view of the node under consideration) mesons $\mathcal{M}$ and baryons $\mathcal{B}$ and $\tilde{\mathcal{B}}$. The quantum moduli space corresponds to the constraint

\beq
\det \mathcal{M}-\mathcal{B}\tilde{\mathcal{B}}=\Lambda^{4M},
\eeq
with $\Lambda$ the dynamical scale of the node under consideration. This equation clearly forces non-zero vevs for the mesons and baryons. The quantum constraint can be efficiently incorporated by introducing a Lagrange multiplier chiral field $X$ and extending the original superpotential $W_0$ to

\beq
W=W_0+X (\det \mathcal{M}-\mathcal{B}\tilde{\mathcal{B}}-\Lambda^{4M}) .
\eeq

Let us now specialize this general discussion for the quiver in \fref{quiver_dP3_to_conifold}.a. Nodes 1 and 4 give rise to mesons $\mathcal{M}$ and $\mathcal{N}$, which can be put in matrix form as follows\footnote{Other conventions for the rows and columns lead to equivalent results.}

\beq
\begin{array}{rcccc}
\mathcal{M} & = & \left(\begin{array}{cc} M_{63} & M_{62} \\ M_{53} & M_{52} \end{array}\right) & = & 
\left(\begin{array}{cc} X_{61} X_{13} & X_{61} X_{12} \\ X_{51} X_{13} & X_{51} X_{12} \end{array}\right) \\ \\
\mathcal{N} & = & \left(\begin{array}{cc} N_{36} & N_{26} \\ N_{35} & N_{25} \end{array}\right) & = &
\left(\begin{array}{cc} X_{34} X_{46} & X_{24} X_{46} \\ X_{34} X_{45} & X_{24} X_{45} \end{array}\right) 
\end{array}
\eeq
In addition, there are also baryons $\mathcal{B}$, $\tilde{\mathcal{B}}$, $\mathcal{C}$ and $\tilde{\mathcal{C}}$ for these two nodes.

Motivated by the structure of the RG cascade we are considering, let us take $\Lambda_1=\Lambda_4\equiv \Lambda$. The deformed geometry is easily recovered along the mesonic branch of the gauge theory, saturating the quantum constraint with meson vevs

\beq
\det \mathcal{M}=\det \mathcal{N}=\Lambda^{4M}.
\eeq

For simplicity, let us consider diagonal vevs for the mesons, i.e. non-zero vevs for $M_{63}$, $M_{52}$, $N_{36}$ and $N_{25}$, while all the others vanish. As a result, nodes 3 and 6 are higgsed down to the diagonal subgroup, which we call 3/6. The same happens for nodes 2 and 5, which are higgsed to a single node 2/5. Carefully studying the gauge theory, it is possible to determine that at low energies it is reduced to the conifold one \cite{Klebanov:1998hh}, whose quiver is shown in \fref{quiver_dP3_to_conifold}.b. We have thus reproduced, via a field theoretic calculation, the results of the geometric analysis.  

\begin{figure}[!ht]
\begin{center}
\includegraphics[width=15.5cm]{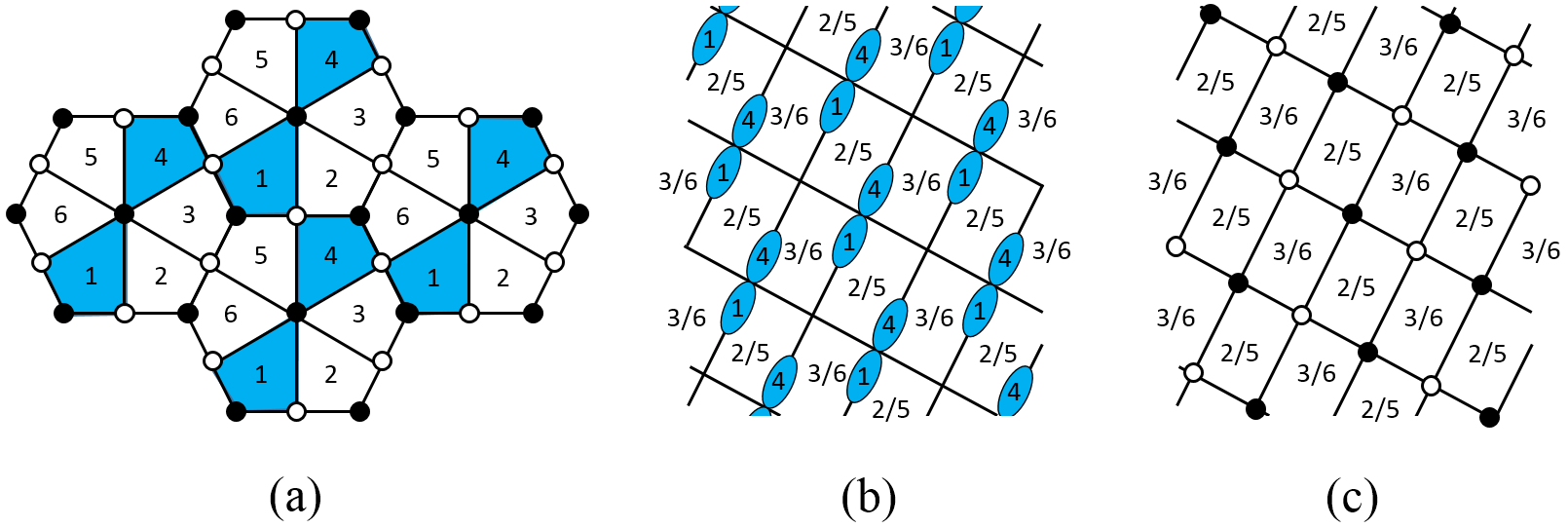}
\caption{a) Dimer for phase 1 of $dP_3$. The steps b) and c) describe the gauge dynamics of the system, resulting in a complex deformation of the moduli space, c.f. \cite{Franco:2005zu,GarciaEtxebarria:2006aq}.}
\label{dimer_deformation_dP3}
\end{center}
\end{figure}

Remarkably, this process can be captured graphically by a transformation of the dimer \cite{Franco:2005zu,GarciaEtxebarria:2006aq}, as shown in \fref{dimer_deformation_dP3}. Shaded faces correspond to the gauge factors whose ranks are changed to account for the introduction of the deformation fractional brane providing the field theory dual of the RR 3-form flux in the warped throat. The remnant dimer diagram describes the gauge theory of D3-branes at the surviving conifold singularity. 

%=======================================================
%=======================================================

\bigskip

%=======================================================
\subsection{Axion Monodromy as Seiberg Duality}
%=======================================================
\label{sec:seiberg}

We are now ready to provide a field theory interpretation of axion monodromy. Anticipating our main conclusion, the bottom line is that the axion monodromy corresponds to a periodic chain of Seiberg dualities whose net effect is to increase the total D3-brane charge in the system. This is triggered by a brane creation effect, in a concrete realization of the mechanism proposed in appendix A.1 of \cite{Marchesano:2014mla}. Although this is very similar to an RG cascade, we emphasize that it is not driven by an RG flow, but rather by the change of a scalar field vev (a would-be modulus, were it not for the monodromy). 
In order to distinguish it from the RG cascade, we refer to the new sequence of dualities induced by rolling of the $\phi$ vev as a {\em monodromy cascade}. 

Let us explain these ideas for the $dP_3$ example. The deformation branes triggering the RG cascade discussed in \sref{section_KT_dP3} correspond to adding $M$ units to the ranks of nodes 1 and 4. As discussed in \sref{section_bottom_throat_dP3}, the RG cascade terminates in a complex deformation generating the finite 3-cycle and is responsible for the exponential suppression of its size. The strong dynamics associated to the deformation is such that the remaining four nodes are higgsed in pairs, 2/4 and 3/5, giving rise to the two gauge groups of the leftover conifold theory. We thus conclude that the $B_2$ field on the surviving $\IS^2$ is associated to the relative gauge coupling of these gauge groups.

\begin{figure}[!ht]
\begin{center}
\includegraphics[width=12.5cm]{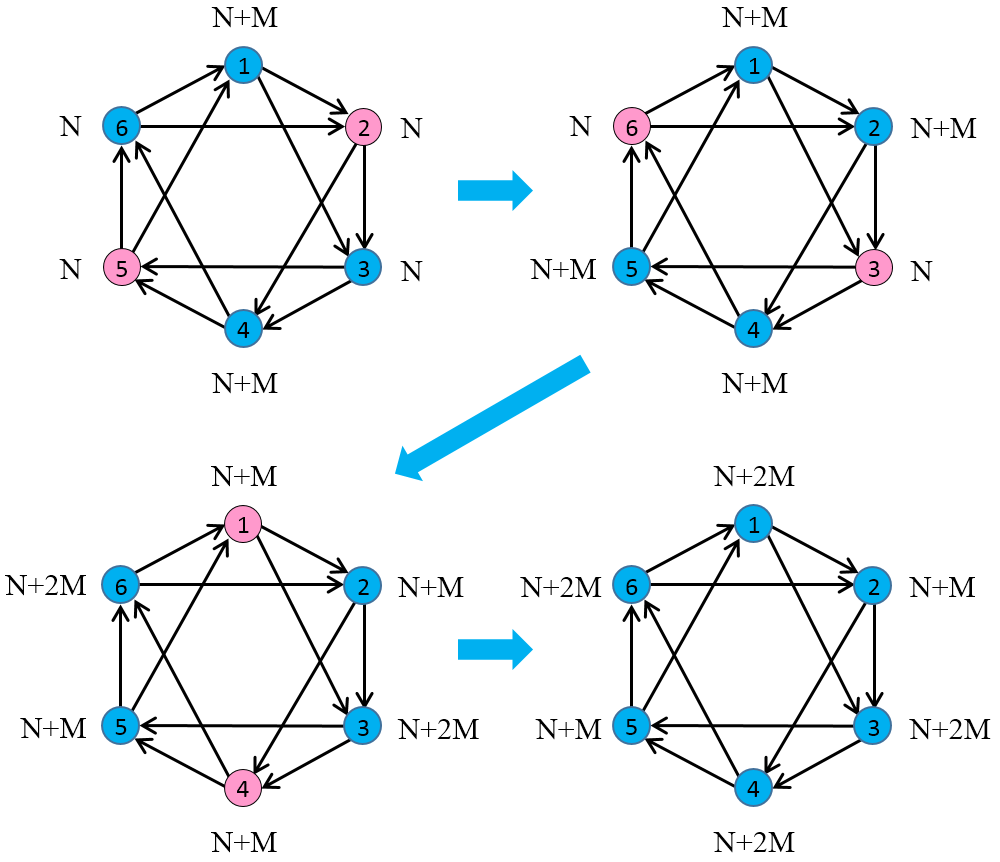}
\caption{A few steps in the monodromy cascade for $dP_3$. The pair of nodes dualized at each time is shown in pink. This monodromy cascade is simply related to RG cascade in \fref{cascade_quiver_dP3} by a 60$^\circ$ rotation in the quiver.}
\label{quiver_monodromy_cascade_dP3}
\end{center}
\end{figure}

It is convenient to organize the six nodes of the $dP_3$ quiver into three pairs: (14), (25) and (36).\footnote{This is an example of a 3-block quiver. We refer the reader to \cite{KN} for a classification of 3-block del Pezzo quivers.} The deformation branes correspond to giving them ranks $(N+M,N,N)$. 
The non-trivial intersection of the 2-cycle associated to $\phi$ and the 3-cycle associated to the flux/fractional branes manifests in the fact that the deformation fractional branes contribute $M$ additional flavors to the node (25) associated to the B-field. Following the analysis in \sref{sec:monodromy}, we expect a non-trivial monodromy for $\phi$ related to the value of $M$. Its realization in the field theory dual is as follows. Changing the $B_2$ field will bring e.g. the node 2/5 past infinite gauge coupling, so we have to Seiberg dualize it (i.e. dualize both 2 and 5 in the $dP_3$ quiver), and the ranks become $(N+M,N+M,N)$. Next, for similar reasons, we dualize 3 and 6, which takes the ranks to $(N+M,N+M,N+2M)$. The next step involves a dualizations of 1  and 4. This is trickier to justify, because the corresponding 2-cycle disappears in the geometric transition, but it is natural to expect that there is a contribution of the $B_2$ field to the gauge couplings of 1 and 4 in the parent geometry. We assume this to be the case, and this will be justified by the correct appearance of the monodromy.\footnote{This can also be justified by the detailed geometry of the fractional branes for all the $dP_3$ gauge groups.} The ranks now become $(N+2M,N+M,N+2M)$. These first steps in the monodromy cascade are shown in \fref{quiver_monodromy_cascade_dP3}. Repeating the same sequence of dualizations on nodes (25), (36) and (14), we obtain once again the original quiver, but with ranks $(N+4M,N+3M,N+3M)$. A period of the monodromy cascade thus involves twelve dualizations and increases the number of D3-branes by $3M$.\footnote{A non-trivial numerical factor, in this case 3, multiplying the number of fractional branes in the change in the number of D3-branes in a cascade period is a generic feature and has been observed in several examples of RG cascades in the literature,  see e.g. \cite{Franco:2004jz,Franco:2005fd}.}

The structure of the monodromy cascade is almost identical to the RG cascade discussed in \sref{section_KT_dP3}. The main difference is that the dualizations are not driven by beta functions. In particular, it is not the gauge groups with higher ranks which are dualized at each step. In fact, the symmetry of this theory is such that the sequence of dualizations in the monodromy cascade is simply related to the RG cascade in \fref{cascade_quiver_dP3} by a 60$^\circ$ rotation in the quiver.

%=======================================================
%=======================================================

Before closing, it is interesting to point out an analogy of the above monodromy with that of the $\theta$ angle in non-supersymmetric gluodynamics \cite{Witten:1998uka} (see also \cite{Dubovsky:2011tu}, and the recent \cite{Dine:2014hwa,Yonekura:2014oja} for applications to axion monodromy inflation). In the latter, there is a periodic coupling in the theory, on which the system has a multi-valued energy dependence. The naive supersymmetric extension, like SYM or SQCD, does not provide a similar monodromy for the $\theta$ angle. In the present case, we start with a conformal theory, with another circle valued marginal operator, which actually controls the gauge coupling (or rather the deviation with respect to to a symmetric value), and which develops a multivalued nature when the system is deformed by the change in ranks.
Hence our construction can be regarded as a natural extension of the $\theta$ angle story to the SUSY setup.

\bigskip

%=======================================================
\section{Type IIA Models}
%=======================================================
%
\label{sec:iia}

The type IIB models in the previous section have the drawback of requiring non-trivial 1-cycles in the compact space, a feature not present in familiar CY compactifications. However, many of the ideas discussed there are valid more generally. As an illustration, we 
consider their implementation in the type IIA setup, where now the required ingredients are 2- and 3-cycles. These IIA models are therefore very amenable to embedding into CY compactifications. Interestingly, the relevant local geometries are again given by the geometric transition studied in the IIB setup, c.f. Appendix \ref{sec:general}.

Consider as a simplified setup a local geometry with two 2-cycles $\Sigma_2$ and $\Pi_2$, and introduce $M$ units of RR 2-form flux $F_2$ on $\Pi_2$. From the general discussion around (\ref{cs-general}), we have the couplings
\beqa
\int_{10d} B_2\wedge F_2 \wedge F_6 \to M\int_{4d} \phi F_4
\eeqa
with 
\beqa
\phi=\int_{\Sigma_2} B_2\; ,\; \int_{\Pi_2}F_2=M\; ,\; F_4=\int_{\Pi'_2}F_6
\eeqa
where $\Pi'_2$ is  such that there is a non-trivial triple intersection number among $\Sigma_2$, $\Pi_2$ and $\Pi_2'$, and we note that the 4d $F_4$ is not the 10d RR 4-form field strength.
Following the general arguments in section \ref{sec:monodromy}, the physical 10d field strength ${\tilde F}_4=dC_3+B_2\wedge F_2$ increases non-trivially along $\Sigma_2\times \Pi_2$ upon shifts of $\phi$. The periodicity structure, leading to the monodromy, is associated to the existence of 4d domain walls changing the value of this flux, given by D4-branes wrapped on $\Pi_2'$. As in \cite{BerasaluceGonzalez:2012zn}, these domain walls are $\IZ_M$-valued and can decay in sets of $M$ by ending on a 4d string given by an NS5-brane wrapped on $\Pi_2\times \Pi_2'$.

The above kind of local models appear naturally in the context of the holographic gauge/gravity duality, as pioneered in \cite{Vafa:2000wi} in the conifold case. Consider a stack of $M$ type IIA D6-branes wrapped on the $\IS^3$ of a deformed conifold, so that below the KK compactification scale they describe pure $SU(N)$ SYM. This string embedding provides a gravity dual, given by type IIA on the resolved conifold, with no D6-branes, and with $M$ units of RR $F_2$ flux over the $\IS^2$. In other words, the gravitational throat solution can be obtained by a geometric transition in which a system of D6-branes on 3-cycles is replaced by $F_2$ flux.
The M-theory lift of the geometric transition has been considered in \cite{Atiyah:2000zz,Atiyah:2001qf}. Notice that in the M-theory setup the $F_2$ flux lifts to the presence of torsion cycles in a Lens space $\IS^3/\IZ_M$, thereby connecting with the description of monodromy by torsion homology in \cite{Marchesano:2014mla}.

Hence, the type IIA picture involves throats which can be constructed with standard toric geometry, and are described by $(p,q)$ web diagrams. Some of the corresponding 2-cycles support RR field strength 2-form fluxes; more specifically, these are the 2-cycles that are traded for a 3-cycle in the geometric transition. Interestingly, this is precisely the reverse process of that exploited in the IIB context in earlier sections.\footnote{We note that in the transition, additional 4-cycles can be blown-up in addition to the 2-cycles; although they in principle could support RR 4-fluxes, and modify the discussion of the CS couplings and the resulting axion monodromy, we skip this possibility since it is unfamiliar from the well-established viewpoint of holographic geometric transitions.}
Concretely, take a toric CY singularity with fractional D5-branes on a (total) 2-cycles class $\Pi_2$ triggering a complex deformation in which a 3-cycle $\Pi_3$ appears supporting RR 3-form flux. One can now use the same geometries in reverse order to engineer a type IIA axion monodromy model as follows. By taking the deformed geometry and wrapping $M$ D6-branes on $\Pi_3$, its strong dynamics triggers a geometric transition in which $\Pi_3$ disappears and there is a RR 2-form flux along the newly created 2-cycle (total class) $\Pi_2$. In the process, there may appear new 2-cycles (and 4-cycles), which support no fluxes and which can be used to produce axion candidates. 

An important difference with the IIB setup is that in general, type IIA throats tend to have several flux-less 2-cycles, and therefore may produce inflation scenarios with multiple fields. This is however a question that must be addressed in specific model building, which is beyond our present scope. 

Another interesting remark is that, as emphasized in \cite{Marchesano:2014mla,McAllister:2014mpa}, type IIA models can produce inflationary models with higher powers in the potential $\phi^p$, $p>2$, for instance by introducing RR 0-form flux (i.e. the Romans mass parameter). It would be interesting to pursue the holographic dual interpretation of this richer scenario.

\bigskip

%=======================================================
\section{Implications for Inflation}
%=======================================================
\label{sec:implications}

One often emphasized requirement of realistic inflationary models in string theory is that they properly address the question of moduli stabilization. Fluxed axion monodromy models have been applied to inflation \cite{Marchesano:2014mla} (see also \cite{Blumenhagen:2014gta,Hebecker:2014eua, Ibanez:2014kia,Arends:2014qca,McAllister:2014mpa} for subsequent work). In this respect, an interesting feature of these models is the intricate relation between the inflationary potential and the moduli stabilization potential,  as emphasized in \cite{Marchesano:2014mla,McAllister:2014mpa}, since the appearance of the monodromy is related to the fluxes stabilizing moduli in the model. When these two flux-induced effects occur at comparable scales, inflation can still proceed, albeit with a flattened potential due to backreaction of the other fields \cite{Dong:2010in}. A more model-independent possibility is to find natural suppression mechanisms for the inflaton sector. This is nicely achieved by our warped throat axion monodromy models.

In our models, the axion arises from the NSNS 2-form field on a 2-cycle localized at the bottom of the throat, so its associated scales are redshifted by the warp factor. 
As usual in fluxed axion monodromy \cite{Marchesano:2014mla}, to lowest order in the canonically normalized inflaton field $\phi$, the action takes this simple chaotic inflation form\footnote{As emphasized in \cite{Marchesano:2014mla,McAllister:2014mpa}, higher power potentials $\phi^p$, $p>2$ can be achieved in 10d models involving additional powers of the NSNS 2-form in the CS couplings.} :
\begin{equation}
S = \int d^4 x \sqrt{-g_4} \left( -\frac{1}{2} (\partial \phi)^2 - \frac{1}{2} \rho^2 \phi^2  + \dots \right)
\end{equation}
where $\rho$ is the physical mass scale, set by the warp factor at the bottom of the throat. Since the full supergravity solution is not available for the throats required for warped axion monodromy,\footnote{In fact, given that the axion $\phi$ is the B-field on a collapsed 2-cycle, its full quantitative most likely requires features beyond the supergravity approximation.} we content ourselves with the estimate, c.f. (\ref{ks-scale})
\beqa
\rho^2\sim M_{\rm UV}^{\, 2}\,  e^{-\frac{4\pi K}{3M g_s}} ,
\eeqa
namely, it is fixed by the size of the complex deformation, which is exponentially suppressed with respect to some bulk/UV scale. A typical value is $M_{\rm UV}\sim \frac{1}{M\alpha'}$ (see e.g. \cite{Kuperstein:2003ih,Frey:2006wv}). The inflation scale admits a nice interpretation in terms of the increase in the $F_5$ flux in the geometry, or in the number of D3-branes in the holographic dual. therefore its contribution scales as
\beqa
V\sim M^2 e^{-\frac{8\pi K}{3M g_s}} T_{\rm D3} .
\eeqa
It is a simple matter to choose parameters of the throat so as to reproduce an inflaton mass of around $10^{13}$ GeV, and an inflation scale of $10^{16}$ GeV, e.g. $K/M\sim 1$, and $g_s\sim 0.1$. These models produce a UV completion of the standard chaotic inflation model, with tensor to scalar ratio of $r\sim 0.13$, in the ballpark of the recent BICEP2 results. 

In order to achieve super-Planckian field ranges, the inflaton axion must wind around its basic period a sufficient number of times. A potential drawback of axions at the bottom of throats is that this period is also redshifted, as follows. An estimate of the axion period (or decay constant) $f_\phi$ is obtained by noticing that upon winding once, the number of flux/D3-branes increases by $qM$, where $q$ is an integer (e.g. $q=3$ in the $dP_3$ example). Using the above values of $V$, $\rho$, the axion period can be estimated to scale as
\beqa
f_\phi\sim  q M e^{-\frac{4\pi K}{3M g_s}} \,M_{\rm UV} .
\eeqa
The period experiences a moderate suppression by the warp factor\footnote{The suppression of the axion periodicity due to warping effects was also observed in \cite{Dasgupta:2008hb}.}, hence comparable with the inflaton mass suppression (e.g. by a factor in the range 100-1000). The period could  be increased by considering larger values of $M$  and $q$  (namely, more involved monodromy cascades, such as those in Appendix \ref{sec:general}), but this would render the models more complicated. Therefore, it is fair to say that the period is typically shortened, and thus the models require many windings of the axion, increased by a factor with is parametrically exponential, but numerically moderate.

This reduced periodicity  leads to an increased packing of the branches of the multivalued potential in a given length of Planck scale field range (by the same factor). One may worry about the stability of super-Planckian field excursions, due to a possible enhancement of the tunneling transitions among branches, unwinding the axion by jumping rather than slow-roll, considered e.g. in \cite{Kaloper:2011jz} (see also \cite{Dubovsky:2011tu}). However, in Appendix \ref{sec:tunnel} we show that the tunneling rates are mainly controlled by the distances in field space, rather than by the branch multiplicity. More concretely, the tunneling is exponentially suppressed by the inverse of the energy difference in the jump, so the closest extra branches contribute in a negligible way, rendering the packing of branches irrelevant. The bottom line is that the exponential suppression overcomes the multiplicative factor from the decay channel multiplicity.
Hence, the extra packing of branches does not introduce additional instabilities. 

Incidentally, it is still possible to realize unsuppressed axion periodicities by realizing the axion as the NSNS 2-form on a 2-cycle collapsed at a curve of singularities, which is supported not only at the bottom of the throat, but also extends radially. These are obtained by complex deformations of singularities, whose remaining singularity is $\IC^2/\IZ_2$, see Appendix \ref{sec:general} for examples. These delocalized axions still have a suppressed mass parameter, since their potential arises from the effect of 3-form fluxes, which are localized at the bottom of the throat (specifically, the increase of the flux (\ref{theflux}) is proportional to $F_3$, which is localized on the $\IS^3$ at the bottom of the throat). On the other hand, their 4d kinetic term involves an integral over the zero mode wavefunction, which spreads radially along the throat and is not suppressed by the warp factor. Therefore their period is not redshifted. There may be additional issues in this scenario, regarding possible backreactions carried out of the throat along the curve of singularities (in analogy with \cite{Conlon:2011qp}). The detailed analysis of this further class of models is however beyond the focus of the present paper.

\medskip

A final interesting observation about our specific realization is that the end of inflation at the minimum $\phi=0$ corresponds to a regime of vanishing B-field at a collapsed conifold 2-cycle. This corresponds to the exotic 4d field theory of tensionless strings considered in \cite{Hanany:1996hq}. It would be interesting to explore the implications of this exotic endpoint for reheating. On top of the theoretical challenge, this would require visible sector model building which is beyond the scope of the present paper, and is left for future investigations.

We conclude this brief discussion with a reference to supersymmetry in our setup. The throats we have considered are supersymmetric, hence the axion is always accompanied by a second scalar partner, the saxion. Having equal mass, the latter can participate non-trivially in the inflationary dynamics and modify the properties of single field scenario. This is certainly an important question for supersymmetric setups, which we hope to address in the future. An alternative is to generalize the construction to the non-supersymmetric setup, possibly by using existing realizations of supersymmetry breaking in throats (see e.g. \cite{Kuperstein:2003ih,Argurio:2006ny,Argurio:2007qk,DeWolfe:2008zy,McGuirk:2009xx}). We leave this discussion for future work.

\bigskip

%=======================================================
\section{Conclusions}
%=======================================================

\label{sec:conclu}

We have introduced warped axion monodromy, a new scenario for inflation in string theory whose main ingredients, axions and warped throats, are commonplace in string compactifications. 

As usual, the underlying shift symmetry of axions, combined with monodromy, results in a protected potential over a super-Planckian inflaton range. In addition, the warped throats can naturally give rise to an exponential hierarchy between the inflation and bulk physics scales. In particular, the framework fits perfectly with standard scenarios of moduli stabilization by fluxes in the bulk. A quadratic, i.e. chaotic, axion potential and its monodromy are generated by couplings of the fluxes that support the throat. 

We discussed implementations of this idea both in type IIA and IIB string theory. In type IIB, we presented explicit constructions based on D-branes on toric CY 3-folds, and investigated them in detail from both a geometric and dual gauge theory viewpoints. The warped throats geometrize RG cascades in the dual gauge theories. On the other hand, the axion parametrizes the motion along what we denoted a monodromy cascade. The creation of D3-branes along this cascade is responsible for the monodromy of the potential. The models can moreover be regarded as gravitational background encoding axion monodromy properties of systems of 5-branes on 2-cycles.

The existence of the monodromy requires the intersection between the 2- and 3-cycles at the bottom of the throat. We reflected on the geometric and gauge theory implications of such intersection.

Finally, we included an appendix with a general classification of an infinite class of geometries giving rise to warped axion monodromy. 

It would be interesting to study the embedding of these throats into global compactifications, as well as determining possible signatures of models based on different throat geometries.

%=======================================================
\section*{Acknowledgments}
%=======================================================

We thank L. Ib\'a\~nez, F. Marchesano and G. Shiu for discussions. A. R. is grateful to the Institute for Particle Physics Phenomenology, Durham University,  for kind hospitality during the course of this work. The work of S. F. and D. G. is supported by the U.K. Science and Technology Facilities Council (STFC). A. R. and A. U. are partially supported by the grants  FPA2012-32828 from the MINECO, the ERC Advanced Grant SPLE under contract ERC-2012-ADG-20120216-320421 and the grant SEV-2012-0249 of the ``Centro de Excelencia Severo Ochoa" Programme. A. R. was also partially supported by the COST Action MP1210.

%\newpage

\bigskip

\appendix

%=======================================================
\section{The Geometry of General Throats}
%=======================================================
\label{sec:general}

%\subsection{Geometry of Complex Deformations}

Geometrizing monodromy inflation in terms of warped throats provides an efficient framework for generating a wide class of generalizations of the simplest models. While exhibiting the same basic behavior at the bottom of the throat, it is natural to expect that the distinctive features of such generalizations might be relevant in concrete applications. For example, they might be important when completing the geometry into a compactification along the lines of \cite{Balasubramanian:2012wd}.

As previously explained, complex deformations of toric singularities translate into decompositions of the corresponding $(p,q)$ webs into sub-webs in equilibrium. In this process, the original external legs are distributed over the sub-webs such that their $(p,q)$ charges add up to zero in each of them. This procedure can be equivalently phrased in terms of the Minkowski sum of toric diagrams \cite{Franco:2005zu}. 

In order to illustrate the flexibility of the throat approach to monodromy inflation, in this appendix we extend our investigation of geometries that result on a conifold after a complex deformation. We will not address the important issue of the intersection between the 2- and 3-cycles which, as mentioned in \sref{sec:seiberg}, is crucial for the axion monodromy to work. In order to find the parent geometry, we start from the $(p,q)$ web for the conifold and add to it different webs. In fact, for the application we are interested in in this article, the problem is strongly constrained and easy to study in full generality. Since the additional webs represent other pieces of the geometry at the IR bottom of the throat, we are interested in them not containing additional 2-cycles. This severely limits the possible webs we can add to the conifold, namely either a line with label $(p,q)$, or a the web associated to $\mathbb{C}^3$, i.e. a set of three semi-infinite lines joining at a vertex. We refrain from a general description of infinite classes of models arising from the second alternative.

\begin{figure}[!ht]
\begin{center}
\includegraphics[width=12cm]{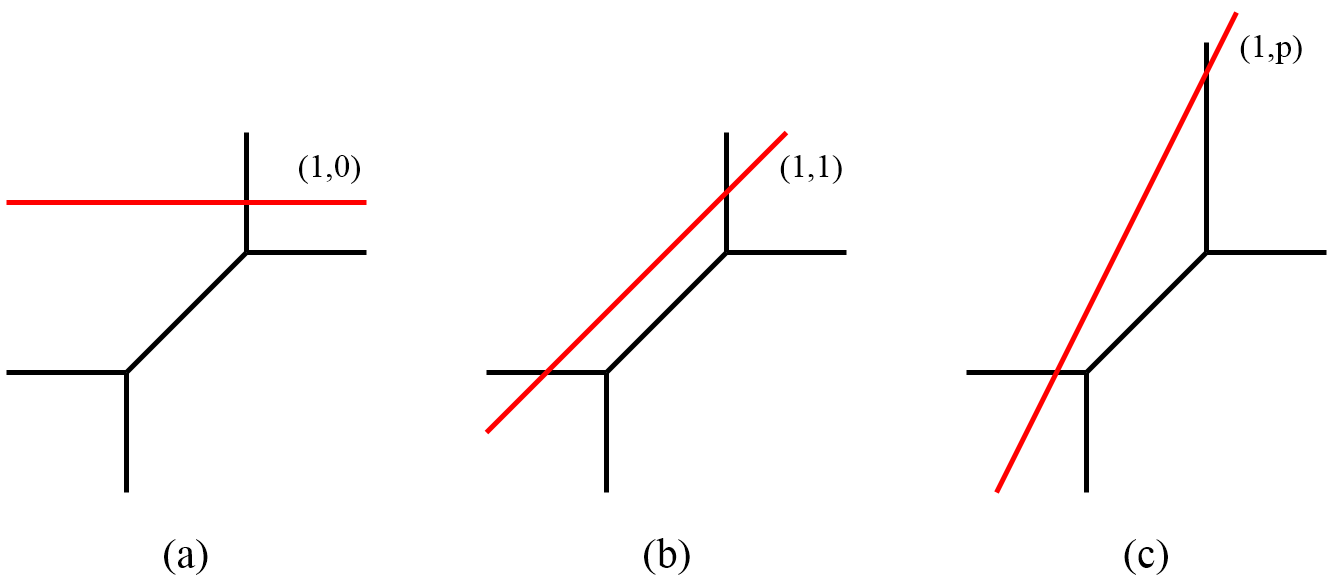}
\caption{Addition of a line with slope $(1,p)$ to the $(p,q)$ web of the conifold. a) $p=0$ gives rise to the conifold/$\mathbb{Z}_2$. b) $p=1$ corresponds to the complex cone over $dP_3$. c) Additional line with general $p$.}
\label{webs_1_line}
\end{center}
\end{figure}

For concreteness, let us focus on adding a line with slope $(p,q) = (1,p)$. In more detail, we add to the web two external legs, with $(p,q)$ charges $(1,p)$ and $(-1,-p)$. This apparently simple setup will turn out to be extremely rich and exhibit interesting dynamics. Thinking in terms of the RG flow in the corresponding gauge theories, we will often refer to the original and leftover geometries as UV and IR geometries. The general configuration is shown in \fref{webs_1_line}. The first two members of this infinite family of geometries, $p=0,1$, correspond to a $\IZ_2$ orbifold of the conifold and the complex cone over $dP_3$ we studied in \sref{section_dP3}.

%=======================================================
%=======================================================

Let us now view the gauge dynamics by which the conifold theory emerges from the corresponding quivers at low energies from some new angles. Let us first consider the $dP_3$ and conifold/$\mathbb{Z}_2$ theories. When going from the IR conifold to the UV $dP_3$, the numbers of 2-cycles and 4-cyles are increased by three and one, respectively. This implies that there are four new independent ways of wrapping D-branes in the singularity. As a result, the parent UV theory has four more gauge groups than the conifold. The confinement plus higgsing process by which the these four additional gauge groups disappear at low energies was explained in detail in \sref{section_bottom_throat_dP3}. The two gauge groups of the conifold are precisely the two diagonal subgroups surviving the Higgs mechanism. In this theory, the higgsings associated to the two confining nodes are actually not independent: the vevs of mesons coming from different nodes trigger the same higgsings.

The situation is rather similar for the conifold/$\mathbb{Z}_2$. When going from the IR conifold to the UV conifold/$\mathbb{Z}_2$, the geometry gets two additional 2-cycles and no new 4-cycle. This implies that the parent theory has two more gauge groups than the conifold one. The disappearance of this pair of gauge groups at low energies is easy to understand: one gauge group confines and another one is lost when two gauge groups are higgsed to the diagonal subgroup by the non-zero vevs in the quantum moduli space. 

The loss of gauge groups in pairs at low energies by a combination of confinement and higgsing leaving behind a diagonal gauge group, as in the previous two theories, is a common feature of many of the explicit examples considered so far in the literature, such as in e.g. \cite{Franco:2005fd}. However, this behavior is not the only possibility and the pattern of higgsings in the IR can be much more elaborate.

\begin{figure}[!ht]
\begin{center}
\includegraphics[width=5cm]{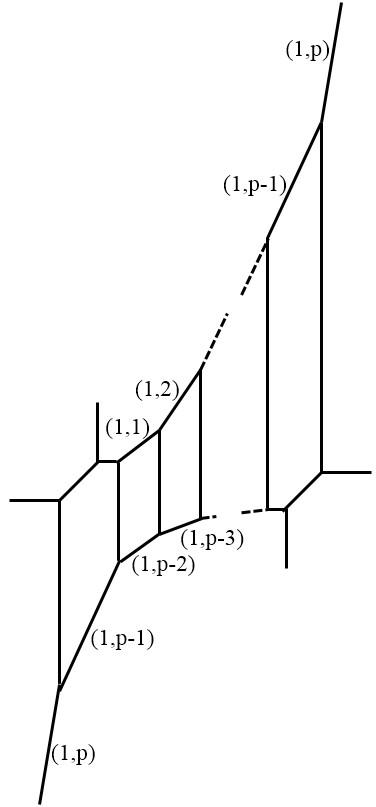}
\caption{The general form of the $(p,q)$ web after adding a line with slope $(1,p)$ to the conifold. Pairs of integers indicate the slopes of internal lines (signs are not important), which have not be drawn to scale.}
\label{general_web}
\end{center}
\end{figure}

Let us now move to the theories for $p>1$, for which the transformation of the geometry when going to the UV becomes more interesting. In this case, the fact that the $(p,q)$ charges of internal branes in the web must be coprime requires an increase in the genus of the web beyond the one that follows from intersecting the new line with the external legs of the conifold. The parent $(p,q)$ web can always be put in the canonical form shown in \fref{general_web}. 

%=======================================================
%=======================================================

\begin{figure}[!ht]
\begin{center}
\includegraphics[width=5.6cm]{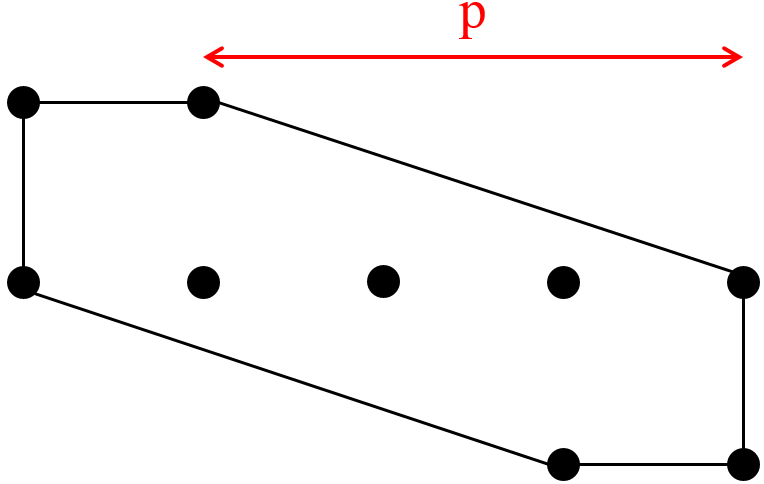}
\caption{Toric diagram for the UV geometry for general values of $p$.}
\label{general_toric_diagram}
\end{center}
\end{figure}

It is also illustrative to look at the corresponding toric diagram, which is presented in \fref{general_toric_diagram}. This general toric diagram clearly becomes the ones for the conifold/$\mathbb{Z}_2$ and the complex cone over $dP_3$ for $p=0$ and $1$, respectively.

%=======================================================
%======================================================= 

From \fref{general_toric_diagram}, we conclude that the parent geometry has $p+2$ 2-cycles and $p$ 4-cycles more than the conifold, accounting for a total of $2p+4$ gauge groups. How does such a large number of gauge groups reduce to two? A more dramatic process than the one for the simple examples mentioned earlier is necessary in order to reduce the number of gauge groups from $2p+4$ down to just two. The generic pattern of confining and higgsed nodes becomes more involved, with higgsed gauge groups forming ``chains". While somehow orthogonal to the main topic of this article, it is interesting to point out that the resulting process is a SUSY, dynamical realization of deconstruction \cite{ArkaniHamed:2001ca}. It would be interesting to investigate what insights can be gained from a geometric implementation of deconstruction such as the one described above.

\begin{figure}[!ht]
\begin{center}
\includegraphics[width=9cm]{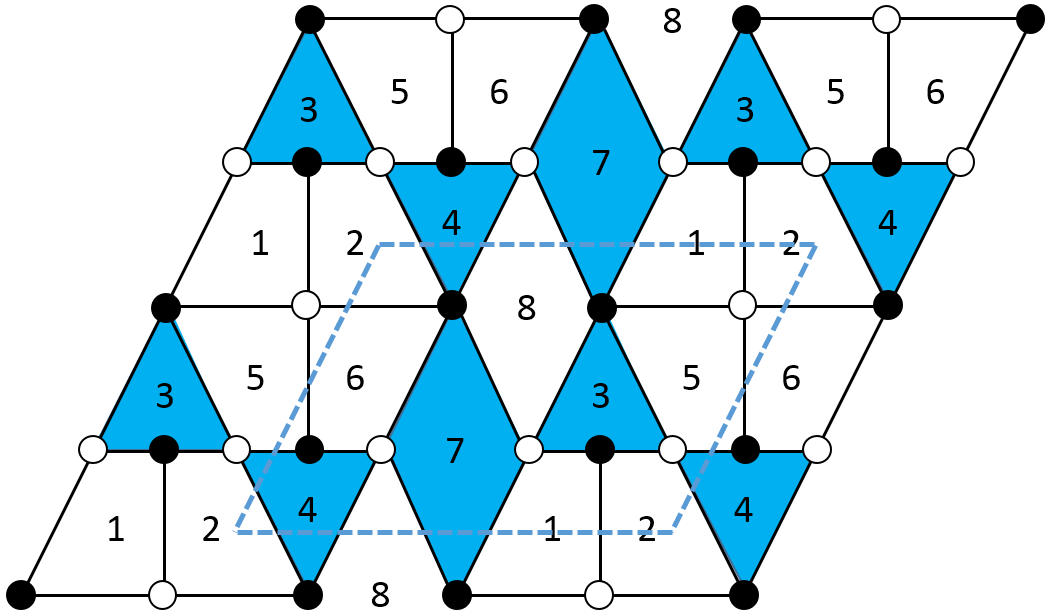}
\caption{Dimer model encoding the gauge theory for $p=2$. It contains 8 gauge groups, 16 chiral fields and 8 superpotential terms. In blue we show the gauge groups associated to the fractional brane giving rise to the deformation.}
\label{p_2_dimer_fractional_branes}
\end{center}
\end{figure}

For illustration, we provide in \fref{p_2_dimer_fractional_branes} the dimer model for the gauge theory associate to $p=2$. We also show the fractional branes responsible for the desired deformation.

%=======================================================
%======================================================= 

\begin{figure}[!ht]
\begin{center}
\includegraphics[width=13cm]{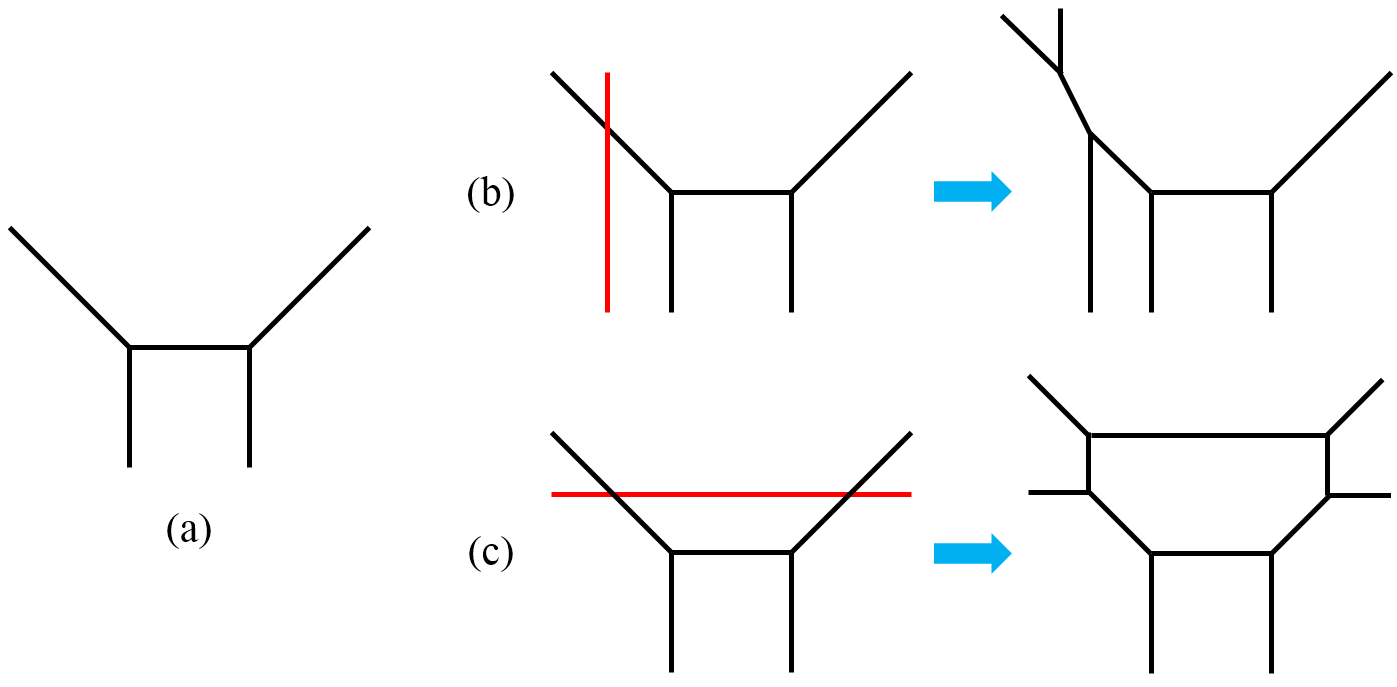}
\caption{The $(p,q)$ webs for $\mathbb{C}^2/\mathbb{Z}_2\times \mathbb{C}$ and for two UV geometries that descend to it after complex deformations.}
\label{C2_Z2_from_deformations}
\end{center}
\end{figure}

Let us conclude this section with some words about other possible generalizations. Since we want the IR geometry to contain a single 2-cycle, we are left with a single possibility other than the conifold, if we restrict to the realm of toric singularities. This geometry is $\mathbb{C}^2/\mathbb{Z}_2\times \mathbb{C}$, whose $(p,q)$ web is shown with some possible UV geometries in \fref{C2_Z2_from_deformations}.  

%=======================================================
%======================================================= 

\begin{figure}[!ht]
\begin{center}
\includegraphics[width=9cm]{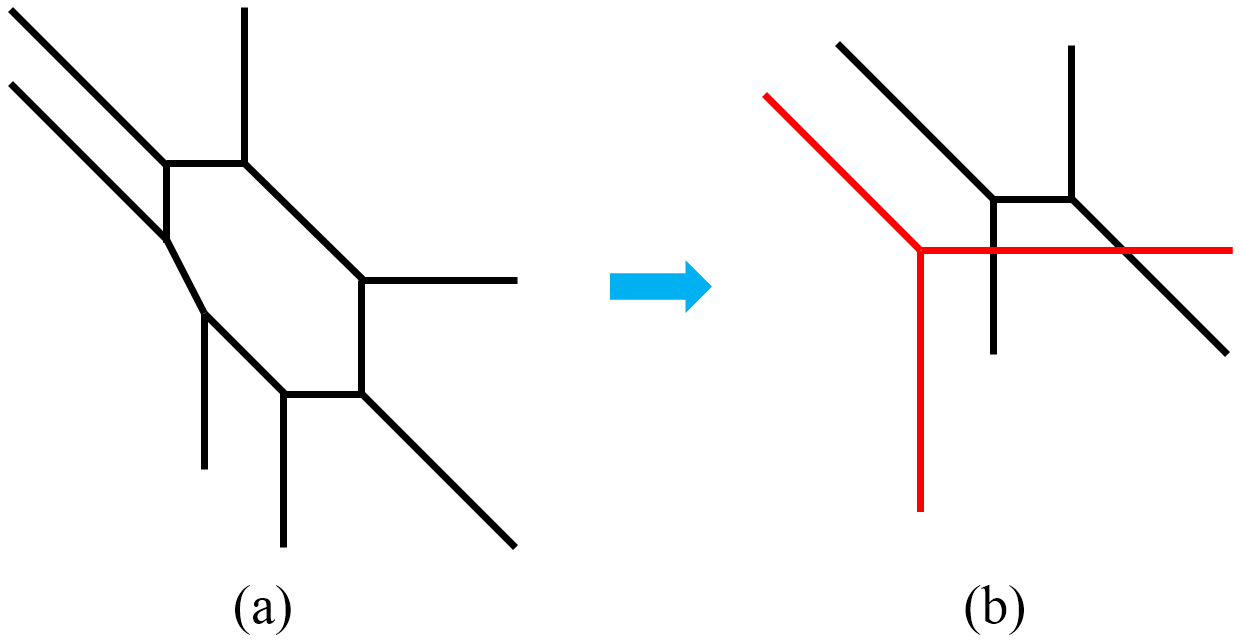}
\caption{The $(p,q)$ webs for $PdP_4$ and its complex deformation to the conifold.}
\label{pdp4}
\end{center}
\end{figure}

Another generalization is when the extra web is described by three semi-infinite lines joining at a vertex (describing a $\IC^3)$. For example,  \fref{pdp4} shows the complex cone over the non-generic del Pezzo surface $PdP_4$. We refrain from a general description of infinite classes of models of this kind.

%=======================================================
%======================================================= 

\bigskip

%=======================================================
\section{No 5d Intersection in an Orbifold of the Conifold}
%=======================================================
\label{sec:noint}

In the general discussion in \sref{sec:monodromy}, we emphasized that the existence of a non-trivial monodromy in the warped throat requires a non-trivial intersecting number between the 2-cycle $\Sigma_2$ and the 3-cycle $\Pi_3$ in the 5d horizon $\IX_5$. As discussed in the $dP_3$ example in \sref{sec:seiberg}, this corresponds to the requirement that the fractional branes contribute to the number of flavors of the nodes associated to the 2-cycle yielding $\phi$. This requirement is generically satisfied, and in particular holds for the infinite class of models in Appendix \ref{sec:general}, except for the case $p=0$. We now describe this case, as an illustration of a warped throat which contains 2- and 3-cycles but does not lead to axion monodromy.

Consider the orbifold of the singular conifold $xy-zw=0$ by the $\IZ_2$ action $x,y\to -x,-y$. By defining invariant monomials $x'=x^2$, $y'=y^2$, the quotient space is described by the geometry $x'y'-z^2w^2=0$. This $\IZ_2$ orbifold of the conifold was discussed in \cite{Uranga:1998vf}, and is also often referred to as the real cone over $L^{2,2,2}$. This is a member of the infinite class of $L^{a,b,c}$ geometries, which were introduced in \cite{Cvetic:2005ft,Martelli:2005wy} and whose gauge theory duals were first found in \cite{Franco:2005sm,Butti:2005sw,Benvenuti:2005ja}. This geometry admits a complex deformation, given by
\beqa
x'y'-z^2w^2=\epsilon \, zw .
\eeqa
It is possible to see that the 3-cycle has $\IS^3$ topology. The deformed geometry still contains a singularity at $x'=y'=z=w=0$, which is a singular conifold. This is manifest when neglecting the higher order $z^2w^2$ term, or equivalently sending the deformation parameter $\epsilon\to \infty$. This can be taken as the singular limit of its small resolution phase, so that there is an $\IS^2$ of vanishing size at that remnant singularity.

\begin{figure}[!ht]
\begin{center}
\includegraphics[width=9cm]{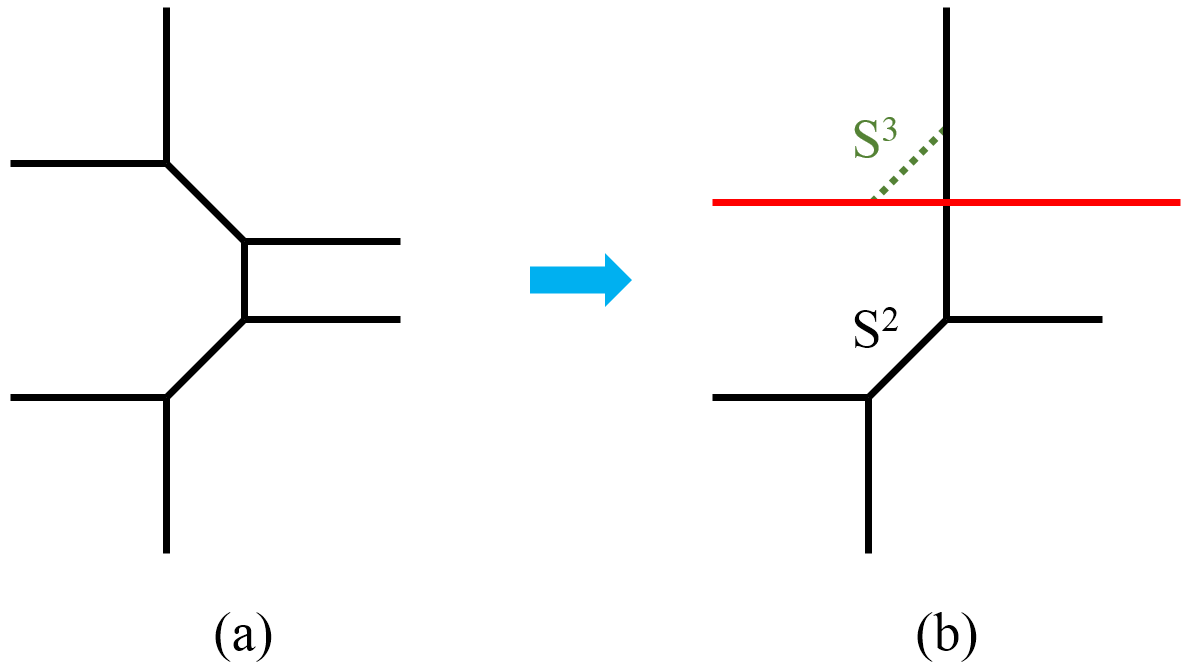}
\caption{\small a) Web diagram of the conifold/$\IZ _2$. For clarity we show the collapsed 2-cycles as slightly blown up. b) Complex deformation showing the 2- and 3-cycles in the resulting geometry.
\label{fig:orbcon}}
\end{center}
\end{figure}

We present the web diagram of the singularity considered in this section and its complex deformation in \fref{fig:orbcon}.

%=======================================================
%=======================================================

\begin{figure}[!ht]
\begin{center}
\includegraphics[width=5cm]{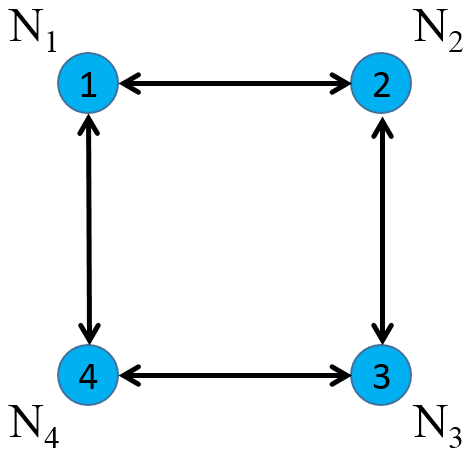}
\caption{Quiver diagram for the obifold/$\mathbb{Z}_2$.}
\label{quiver_orbifold_conifold}
\end{center}
\end{figure}

\begin{figure}[!ht]
\begin{center}
\includegraphics[width=15cm]{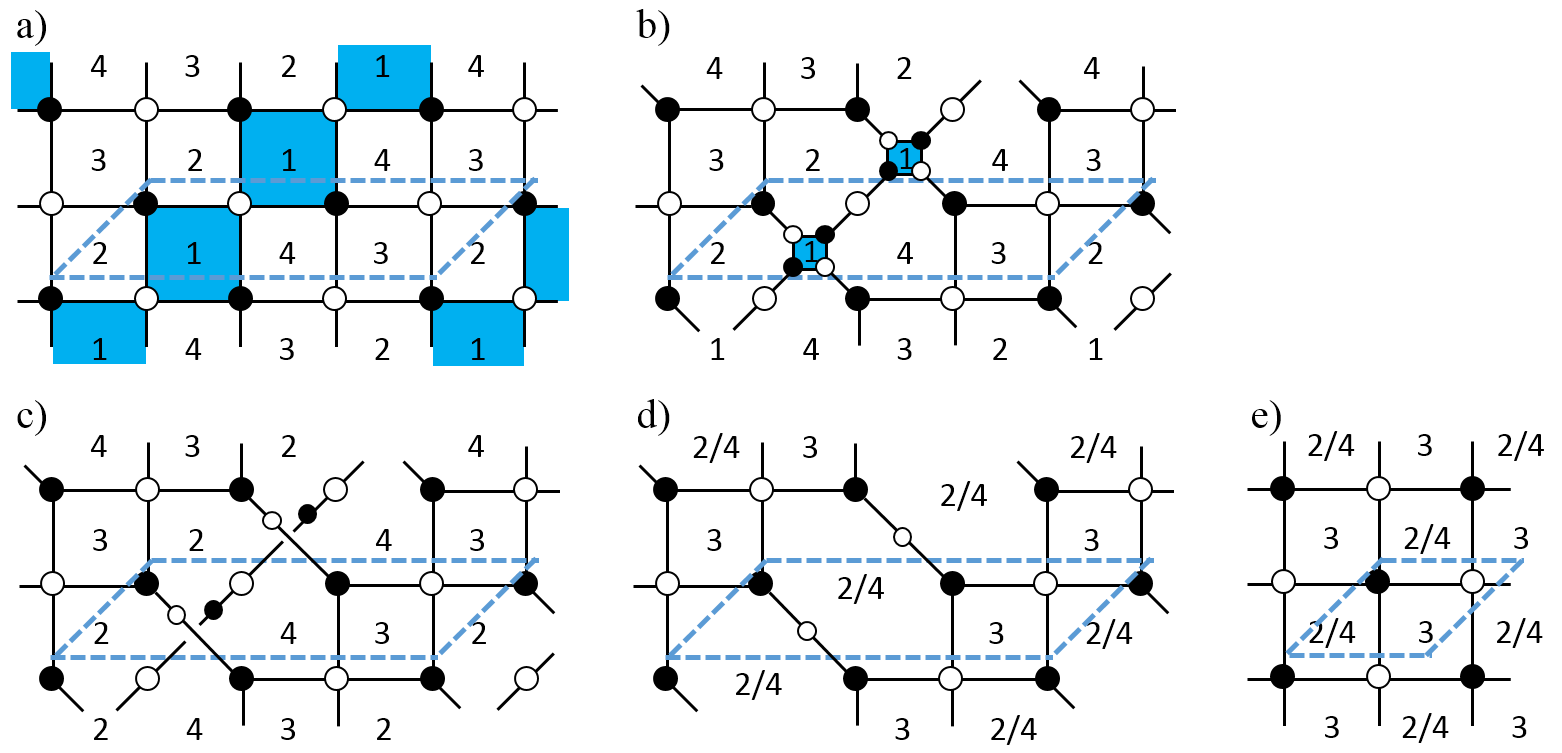}
\caption{\small a) Dimer of the orbifold of the conifold. The steps b)-e) describe the gauge dynamics of the system, resulting in a complex deformation of the moduli space, c.f. \cite{GarciaEtxebarria:2006aq}.
\label{fig:conidef2}}
\end{center}
\end{figure}

Let us now turn to describing the holographic dual field theory, and its description of the complex deformation.
The underlying field theory corresponds to the quiver diagram in \fref{quiver_orbifold_conifold}, with superpotential

\beq
W= \tr( X_{12}X_{21}X_{14}X_{42}-X_{21}X_{12}X_{23}X_{32}+X_{32}X_{23}X_{34}X_{43}-X_{43}X_{34}X_{41}X_{14}) .
\eeq
The theory is nicely encoded by the dimer diagram in \fref{fig:conidef2}.a. 

%=======================================================
%=======================================================

%=======================================================
%=======================================================

The throat we are interested in corresponds to introducing $M$ D5-branes on the 2-cycle associated to e.g. node 1, namely $N_1=N+M$, $N_2=N_3=N_4=N$. As in \cite{Klebanov:2000hb}, this theory has a non-trivial RG flow along which a cascade of Seiberg dualities effectively reduces the value of $N$, while keeping the fractional branes intact. The IR dynamics corresponds effectively to $N_1=2M$, $N_2=N_3=N_4=M$, so that the strong dynamics at node 1 (which has $N_f=N_c$) produces a complex deformation of the moduli space, matching that of the geometry. The field theory analysis (we refer the reader to \cite{Franco:2005fd} for a thorough discussion and several explicit examples) can be mapped a to simple diagrammatic process in the dimer diagram, shown in \fref{fig:conidef2}. The result agrees with the fact that the deformed geometry is not completely smooth but contains  a remnant conifold singularity. 

\begin{figure}[!ht]
\begin{center}
\includegraphics[width=15cm]{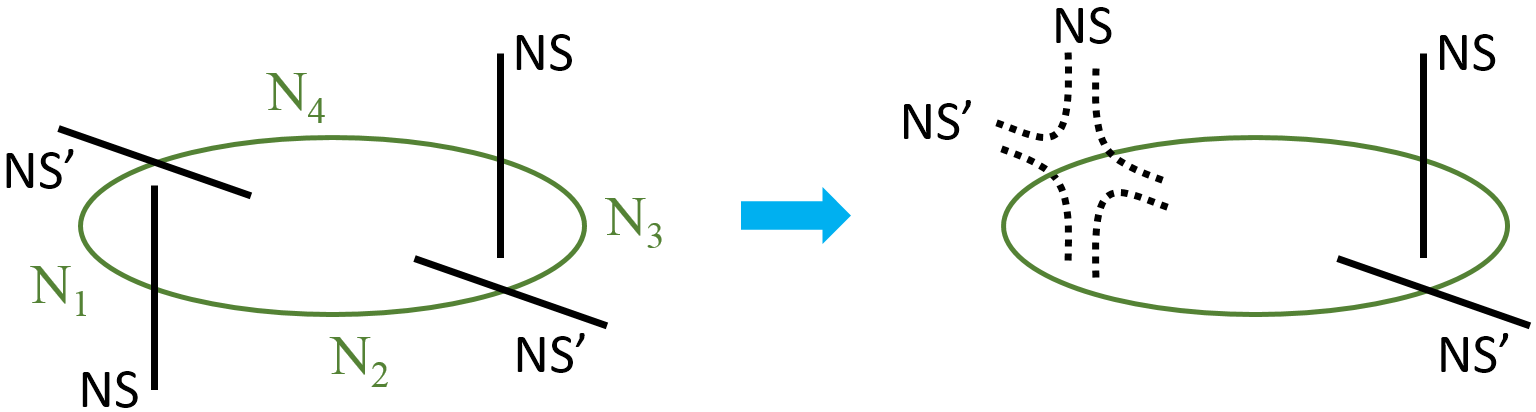}
\caption{\small a) Hanany-Witten T-dual picture of the system of D3-branes at the orbifold of the conifold. It involves D4-branes suspended in one direction between relatively rotated NS5-branes (denoted by NS- and NS'-branes). b) The geometric transition is described as the recombination of one NS- and one NS'-brane, which effectively disappear from the picture. The remaining NS- and NS'-brane describe the T-dual of the remaining conifold singularity.}.
\label{fig:hw}
\end{center}
\end{figure}

A complementary description of the systems is in terms of a Hanany-Witten system \cite{Hanany:1996ie}, which is related to the D3-brane system by a T-duality (along the  $U(1)$ orbit of the action $x' \to e^{i\alpha} x'$, $y'\to e^{-i\alpha} y'$) \cite{Uranga:1998vf}, see Figure \ref{fig:hw}.a. It contains D4-branes along 0123 and streched along 6 between two NS-branes (along 012345) and two rotated NS-branes (along 012389, henceforth denoted NS'-branes). The direction 6 is the T-dual circle. The $\IZ_2$ orbifold point in moduli space corresponds to the ordering of NS-branes as NS-NS'-NS-NS'. The D4-branes suspended in each of the intervals in 6 corresponds to a gauge factor, and the gauge group, matter content and superpotential can be obtained using standard rules (see \cite{Uranga:1998vf}, and \cite{Giveon:1998sr} for a review), and shown to agree with the above dimer. 

%=======================================================
%=======================================================

The complex deformation is attained by suspending $M$ D4-branes on e.g. interval 1, in addition to the  $N$ D4-branes in all intervals. The RG duality cascade corresponds to the dynamical motion of the NS and NS'-branes bounding the initial interval 1, moving them across other branes, which results in an effective change in the value of $N$ due to brane creation effects. The infrared configuration only contains the $M$ D4-branes on interval 1, while others are empty, and the strong dynamics maps to the recombination of the NS- and NS'-branes, see Figure \ref{fig:hw}.b. This heuristic description follows from the M-theory lift of the configuration, as in \cite{Witten:1997ep}. The $\IS^3$ in the complex deformation is mapped to a 1-cycle (or rather, the disk bounded by it) in the recombined 5-brane, while the remnant conifold singularity is T-dual to the remnant system of one NS and one NS'-brane.

We can now see that there is no axion monodromy in the present setup. The field $\phi$ (the B-field on the conifold 2-cycle) maps to the modulus controlling the distance between the NS- and NS'-branes in the HW picture. Moving the axion around its period corresponds to moving e.g. the NS'-brane around the circle. Even accounting for brane creation effects, it is easy to check that the resulting configuration is exactly identical to the original one, and there is no net creation of D3-branes and no monodromy. Both in the HW or the dimer pictures, the hint is that the gauge factor 3  associated to the 2-cycle gets no flavors from the fractional brane inducing the deformation. This is the field theory counterpart of the 2-cycle $\Sigma_2$ and the 3-cycle $\Pi_3$ not having intersection in $\IX_5$.

It is tempting to blame the failure to generate monodromy to the presence of parallel NS5-branes in the HW picture, or more precisely of parallel external legs in the two $(p,q)$ sub-web diagrams in the complex deformation, c.f. \fref{fig:orbcon}. This is however not the origin of the problem, since it is easy to construct examples with such parallel external legs and with a nice monodromy. For instance, the cone over $PdP_4$ deformed to a conifold, c.f. \fref{pdp4}, has parallel legs, but also monodromy (in fact, related to that of $dP_3$ by a simple higgssing process). Empirically, the crucial difference between the orbifold of the conifold and the examples with actual monodromy (like $dP_3$, $PdP_4$ or the models $p\neq 0$ in Appendix \ref{sec:general}) seems to be that the sub-webs of the former intersect only once, whereas the latter have sub-webs with multiple intersections.

\bigskip
 
%=======================================================
\section{Tunneling Between Branches}
%=======================================================
\label{sec:tunnel}

In models of axion monodromy inflation the inflaton potential is multivalued. In our warped realizations, the decay constant is suppressed and, as a result, the number of windings becomes equally increased. It thus becomes important to address the possibility of the axion tunneling to lower branches, which would spoil large-field inflation. Following \cite{Kaloper:2011jz}, here we estimate the tunneling probability based on standard thin-wall approximation \cite{Kobzarev:1974cp,Coleman:1977py,Callan:1977pt}. We will find that the tunneling probability to lower branches is indeed highly suppressed. We shall first review the necessary tools for the computation and then apply them to our particular setup.

\bigskip

%=======================================================
\subsection{The Thin-Wall Approximation}
%=======================================================

Starting from a field which is initially in a false vacuum state $\phi_i$ over the entire space, it is possible to nucleate a bubble by tunneling to a lower energy vacuum state $\phi_f$. The region interpolating between both regions is known as the {\it wall} and is characterized by its tension $\sigma$. The probability for the tunneling to happen goes as
\begin{equation}
P \sim \exp \left( - \Delta S_E \right) \quad, \quad \Delta S_E = S_E (\phi_f ) - S_E (\phi_i )
\end{equation}
where $S_E$ is the Euclidean action of the corresponding field configuration. Once the bubble has nucleated, the space is divided into three regions: the interior, the exterior and the wall of the bubble. In the thin-wall approximation, the various contributions to $\Delta S_E$ are given by:
\begin{equation}
\Delta S_{E, \text{int} } \sim - \Delta V r^4 \ \ \ , \ \ \ \Delta S_{E, \text{wall}} \sim \sigma r^3 \ \ \ ,\ \ \ \Delta S_{E, \text{ext}} = 0
\end{equation}
where $\Delta V \equiv V(\phi_i ) - V(\phi_f )>0$ and $r$ is the bubble radius. The radius is fixed by maximizing the probability for the bubble to form and is such that the energy released within the bubble goes to the wall, giving rise to its tension $\sigma$.

\bigskip

%=======================================================
\subsection{Tunneling Probability}
%=======================================================

\begin{figure}[!h]
\begin{center}
\includegraphics[width=15cm]{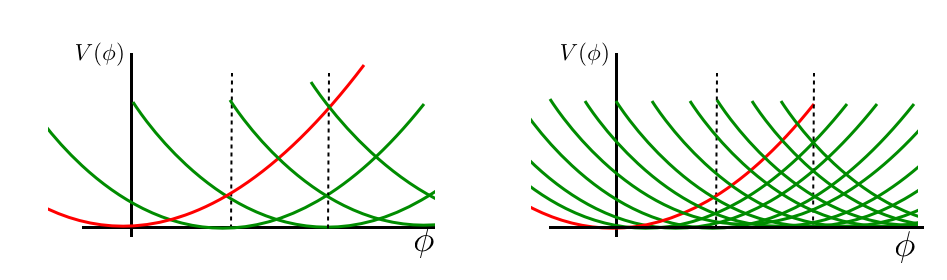}
\caption{\small The different branches of the potential (dotted lines indicate the value of the Planck scale). Warping decreases the decay constant from some value on the top of the throat (left) to a lower value at its bottom (right). This packing of branches implies that for some initial state there will always be a branch of very small energy and same field value. It also increases the amount of possible tunneling routes.}
\label{fig:pack}
\end{center}
\end{figure}

In our type IIB setup the tunneling corresponds to the generation of D3-branes that extend along the radial direction of the throat, taking the energy from the interior of the 3-sphere wall and shifting $F_5$ by an integer amount. Despite the more intricate details of this particular construction, such as the fact that this is a 10d theory with gravity and the tunneling happens between different branches of the potential, the thin-wall analysis can be applied to this scenario to estimate the tunneling probability. In this approximation, we obtain \cite{Kaloper:2011jz}
\begin{equation}
P \sim \text{exp}\left(- \frac{27 \pi^2 \sigma^4}{2 (\Delta V)^3} \right) .
\end{equation}
Here we have assumed that the tunneling is such that $\phi_i = \phi_f$; processes for which $\phi_i \neq \phi_f$ have an additional exponential suppression. We are now equipped to tackle the tunneling to multiple branches. First, we see that the most likely jumps, i.e.\ those with least Euclidean action, are those with a large energy difference. The dominating tunneling process will hence be from the highest-energy state, i.e.\ at the start of inflation, to the branch of lowest energy with the same field value. In fact, as it is from \fref{fig:pack}, a simple estimate can be obtained by approximating the potential at this lowest branch to be zero, giving $\Delta V \simeq V_i$.
%=======================================================
%=======================================================

Since the tension $\sigma\sim f_{\phi}/\alpha'$ we conclude the most likely tunneling process is heavily suppressed
\begin{equation} \label{eq:maxprobability}
P_{\text{max}} \sim \text{exp}\left(- \frac{27 \pi^2 \sigma^4}{2 V_i^3} \right) \ll 1 .
\end{equation}
The lifetime for tunneling to the lowest branch is thus much longer than the time scale of inflation.

Finally, we must take into account the large number possible tunneling channels due to the packing of multiple branches due to warping, as shown in \fref{fig:pack}.  This multiplicity turns out not to be dangerous for our model, since it is overcome by the additional exponential suppression of nearby branches. 

\bigskip

%=======================================================
%=======================================================

\end{document}